\documentclass[useAMS,usenatbib]{mn2e}

\usepackage{aas_macros}

\usepackage{graphicx}
\graphicspath{{./figures/}}

\usepackage{amsmath}
\usepackage{color}

\usepackage{units}

\usepackage{multirow}

\newcommand{\nuclei}[2]{\ensuremath{\mathrm{^{#1}#2}}}

\newcommand{\carbon}[1][12]{\nuclei{#1}{C}}

\newcommand{\oxygen}[1][16]{\nuclei{#1}{O}}
\newcommand{\fluorine}[1][19]{\nuclei{#1}{F}}
\newcommand{\neon}[1][20]{\nuclei{#1}{Ne}}
\newcommand{\sodium}[1][23]{\nuclei{#1}{Na}}
\newcommand{\magnesium}[1][24]{\nuclei{#1}{Mg}}

\newcommand{\msun}{M_\odot}
\newcommand{\msunyr}{M_\odot\,\mathrm{yr}^{-1}}

\newcommand{\revision}{Revision b07f982950e1c5e144321cabb4d3e80735c26383}

\title[Evolution of ONeMg Cores]{Thermal Runaway During the Evolution of ONeMg Cores towards Accretion-Induced Collapse}
\author[Schwab et al.]{
Josiah Schwab$^{1,2}$,
Eliot Quataert$^{1,2}$,
Lars Bildsten$^{3,4}$
\\
$^1${Physics Department, University of California,
     Berkeley, CA 94720, USA}  \\
$^2${Astronomy Department and Theoretical Astrophysics
     Center, University of California, Berkeley, CA 94720, USA} \\
$^3${Department of Physics, University of California, Santa Barbara, CA 93106} \\
$^4${Kavli Institute for Theoretical Physics, Santa Barbara, CA 93106} \\
}

\begin{document}
\date{\revision}

\maketitle

\begin{abstract}
  We study the evolution of degenerate electron cores primarily composed
  of the carbon burning products $\oxygen$, $\neon$, and $\magnesium$
  (hereafter ONeMg cores) that are undergoing compression. Electron
  capture reactions on $A=20$ and $A=24$ isotopes reduce the electron
  fraction and heat the core. We develop and use a new capability of the
  Modules for Experiments in Stellar Astrophysics (MESA) stellar
  evolution code that provides a highly accurate implementation of these
  key reactions. These new accurate rates and the ability of MESA to
  perform extremely small spatial zoning demonstrates a thermal runaway
  in the core triggered by the temperature and density sensitivity of
  the $^{20}$Ne electron capture reactions. Both analytics and numerics
  show that this thermal runaway does not trigger core convection, but
  rather leads to a centrally concentrated ($r<{\rm km}$) thermal
  runaway that will subsequently launch an oxygen deflagration wave from
  the center of the star.  We use MESA to perform a parameter study that
  quantifies the influence of the $\magnesium$ mass fraction, the
  central temperature, the compression rate, and uncertainties in the
  electron capture reaction rates on the ONeMg core evolution.
  This allows us to establish a lower limit on the central density at
  which the oxygen deflagration wave initiates of $\rho_c \ga
  \unit[8.5\times10^9]{g\,cm^{-3}}$.  Based on previous work and
  order-of-magnitude calculations, we expect objects which ignite oxygen
  at or above these densities to collapse and form a neutron star.
  Calculations such as these are an important step in producing more
  realistic progenitor models for studies of the signature of
  accretion-induced collapse.
\end{abstract}


\section{Introduction}
\label{sec:intro}

In this paper, we study the evolution of degenerate electron cores primarily
composed of the carbon burning products $\oxygen$, $\neon$, and
$\magnesium$ which are undergoing compression.  Such objects can arise
in several contexts: the late stages of evolution for super asymptotic
giant branch (SAGB) stars
\citep[e.g.][]{Miyaji87,Ritossa99,Takahashi13,Jones13}, where the compression
is caused by the deposition of material from exterior shell-burning;
in a binary system with a massive ONeMg white dwarf (WD)
\citep[e.g.][]{Nomoto91}, where the compression is caused by accretion
from a non-degenerate companion; or as the remnant of a WD-WD merger,
where the compression is caused by the cooling of the outer layers
\citep[e.g.][]{Saio85}.

As the core is compressed, the electron Fermi energy rises, eventually
triggering exothermic electron capture reactions.  Typically,
exothermic captures on $\neon[20]$ release enough energy to cause
thermonuclear ignition of $\oxygen[16]$ and formation of a
deflagration.  The final fate of the core (either explosion or
collapse) is determined by a competition between the energy release
from the outgoing oxygen deflagration and the energy losses and
decline in the electron fraction due to electron captures on the
post-deflagration material, which has burned to nuclear statistical
equilibrium (NSE).  The evolution of these cores has been a subject of
considerable previous study
\citep[e.g.][]{Miyaji80,Nomoto84a,Isern91, Canal92, Gutierrez96, Gutierrez05, Jones14}.

However, we revisit this topic (i) to test the effect of using the
state-of-the-art Modules for Experiments in Stellar Astrophysics
(MESA) stellar evolution code \citep{Paxton11,Paxton13,Paxton15}, (ii) to
demonstrate the effects of using the latest nuclear reaction rates
\citep{MartinezPinedo14}, (iii) to perform a more detailed parameter
study of the effects of a number of quantities, including the
accretion rate $\dot{M}$, magnesium mass fraction $X_{\mathrm{Mg}}$,
and initial core temperature, $T_c$, and (iv) to provide analytic
estimates of the evolution up-to and including the onset of the oxygen
deflagration.

In the present paper, we follow the common treatment in the
literature and parameterize the evolution of ONeMg WDs as they
approach the Chandrasekhar mass via compression of the outer layers.
In future work we will assess whether the revised evolutionary model
of WD merger remnants proposed by \citet{Shen12} and \citet{Schwab12}
modifies the likelihood of AIC in super-Chandrasekhar WD mergers. In
\S~\ref{sec:mesa} we describe the treatment of weak reactions in the
MESA code.  In \S~\ref{sec:analytics} we provide analytic estimates
relevant to the evolution of the core.  In
\S~\ref{sec:mesa-calculations} we discuss the inputs to our MESA
calculations and in \S~\ref{sec:parameters} present the results of
these numerical simulations.  \S~\ref{sec:discussion} discusses the
final fate of these cores.  In \S~\ref{sec:conclusions} we draw our
conclusions and describe some important avenues for future work.

\section{Weak Reactions in MESA}
\label{sec:mesa}

Weak reactions, specifically electron-capture and beta-decay, are
central to the evolution of accreting degenerate ONeMg cores.  The
reduction in electron fraction (and corresponding reduction in
pressure) due to electron captures accelerates the contraction of the
cores and the entropy generation from these electron captures can
directly ignite thermonuclear reactions.

This study makes use of MESA, a state-of-the-art open source code for
stellar evolution calculations \citep{Paxton11,Paxton13}.  In
particular we use the capability to calculate weak reaction rates
directly from nuclear level and transition data, which is documented
in the upcoming MESA Instrument Paper III \citep{Paxton15}.  This
section summarizes the input data to this capability.  The precise
expressions which are evaluated as part of MESA's on-the-fly weak
reaction treatment are given in Appendix~\ref{sec:ecapture}.

We restrict ourselves to considering only a small set of $A=24$
isotopes ($\magnesium[24]$, $\sodium[24]$, $\neon[24]$) and $A=20$
isotopes ($\neon[20]$, $\fluorine[20]$, $\oxygen[20]$).  Over the
range of thermodynamic conditions encountered during the evolution of
ONeMg cores, roughly $9 \la \log_{10} \rho \la 10$ and
$8 \la \log_{10} T \la 9$ (in cgs units), \citet{Takahara89}
identified the transitions that dominate the rate of each reaction.
We consider only this limited set of transitions; they are listed in
Table~\ref{tab:transitions}.  We have taken the comparative half-lives
of these reactions from the up-to-date information compiled in
\citet{MartinezPinedo14}.

\begin{figure}
  \centering
  \includegraphics[width=\columnwidth]{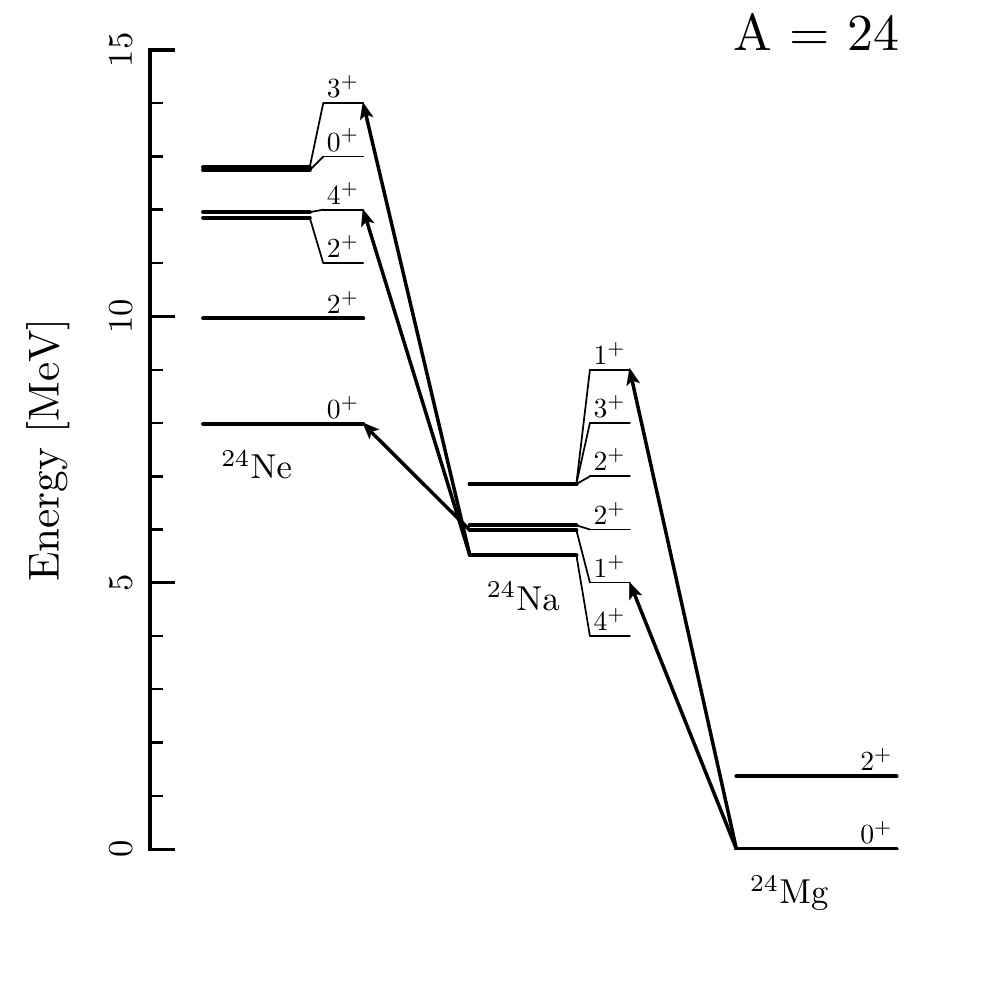}
  \caption{Energy level diagram for the $A=24$ nuclei that we
    consider.  The $J^{\pi}$ values are sometimes given an arbitrary
    offset (indicated via thin lines) in order to enhance
    legibility. The transitions we consider are indicated with
    arrows.}
  \label{fig:levels_MgNaNe}
\end{figure}

\begin{figure}
  \centering
  \includegraphics[width=\columnwidth]{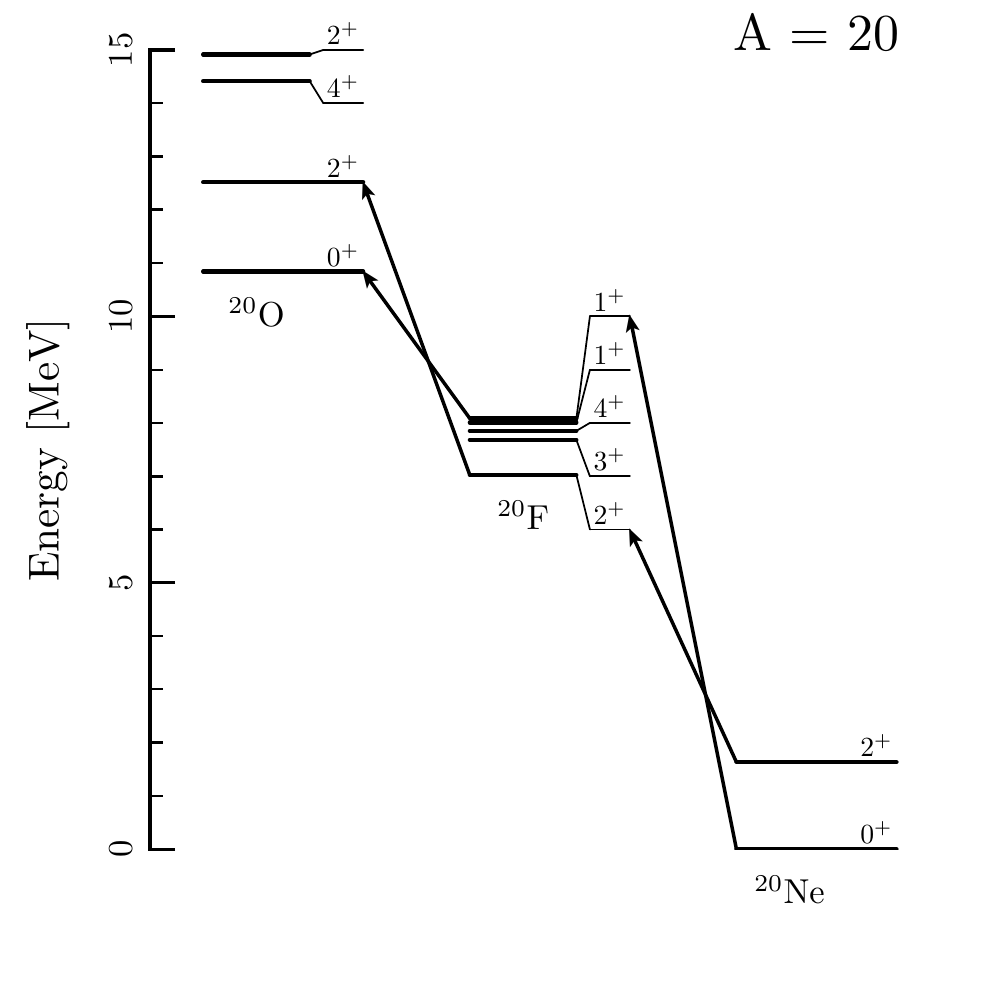}
  \caption{Energy level diagram for the $A=20$ nuclei that we
    consider.  The $J^{\pi}$ values are sometimes given an arbitrary
    offset (indicated via thin lines) in order to enhance
    legibility. The transitions we consider are indicated with
    arrows.}
  \label{fig:levels_NeFO}
\end{figure}

In order to more easily visualize the data in
Table~\ref{tab:transitions}, we present energy level diagrams for the
$A=24$ (Fig.~\ref{fig:levels_MgNaNe}) and $A=20$
(Fig.~\ref{fig:levels_NeFO}) nuclei.  These figures are modeled after
those found in \citet{Takahara89}. The level structure of these nuclei
is drawn from recent compilations of nuclear data
\citep{Tilley98,Firestone07b}.  We show all of the low-lying states
that we consider, labeled with their $J^{\pi}$
($\mathrm{spin}^\mathrm{parity}$) values.  The arrows indicate the
limited set of transitions that we consider, which are only those
which are ``allowed'' (Gamow-Teller: $J_i = J_j, J_j \pm 1$,
$\pi_i \pi_j = 1$; excluding $J_i = J_f = 0$).

\begin{table}
  \begin{tabular}{llllllll}
    \hline
    Initial & Final & $Q_\textrm{g}$ & $E_\mathrm{i}$ & $J_\mathrm{i}^\pi$ & $E_\mathrm{f}$  & $J_\mathrm{f}^{\pi}$ & $\log_{10}(ft)$ \\
    \hline
    \multirow{2}{*}{$\magnesium[24]$} & \multirow{2}{*}{$\sodium[24]$} & \multirow{2}{*}{$5.515$} &
                                                                                                    0.000 & $0^+$ & 0.472 & $1^+$ & 4.815 \\
            & & & 0.000 & $0^+$ & 1.347 & $1^+$ & 3.838 \\
    \hline
    \multirow{3}{*}{$\sodium[24]$} & \multirow{3}{*}{$\neon[24]$} & \multirow{3}{*}{$2.467$} &
                                                                                               0.000 & $4^+$ & 3.972 & $4^+$ & 6.209 \\
            & & & 0.000 & $4^+$ & 4.866 & $3^+$ & 4.423 \\
            & & & 0.472 & $1^+$ & 0.000 & $0^+$ & 4.829 \\
    \hline
    \multirow{3}{*}{$\neon[20]$} & \multirow{3}{*}{$\fluorine[20]$} & \multirow{3}{*}{$7.025$} &
                                                                                               1.634 & $2^+$ & 0.000 & $2^+$ & 4.970 \\
            & & & 0.000 & $0^+$ & 1.057 & $1^+$ & 4.380 \\
            & & & 0.000 & $0^+$ & 0.000 & $2^+$ & \textit{9.801} \\
    \hline
    \multirow{2}{*}{$\fluorine[20]$} & \multirow{2}{*}{$\oxygen[20]$} & \multirow{2}{*}{$3.815$} &
                                                                                                   0.000 & $2^+$ & 1.674 & $2^+$ & 5.429 \\
            & & & 1.057 & $1^+$ & 0.000 & $0^+$ & 4.211 \\
    \hline
  \end{tabular}

  \caption{The transitions used in the rate calculations.  They are
    written as electron capture transitions, but the same transitions were
    used for beta-decay (swapping initial and final states).  $Q_g$ is the
    energy difference between the ground states of the isotopes.
    $E_\mathrm{i}$ and $E_\mathrm{f}$ are respectively the excitation
    energies of the initial and final states, relative to the ground
    state.  $J^\pi_\mathrm{i}$ and $J^\pi_\mathrm{f}$ are the spins and
    parities of the initial and final states.  Allowed transitions do not
    have parity changes. $(ft)$ is the comparative half-life in seconds,
    taken from \citet{MartinezPinedo14} by dividing the constant
    \unit[6144]{s} by their tabulated values of the transition matrix
    elements.  The italicized $(ft)$ value indicates an experimental upper
    limit; the effects of this transition will be discussed in
    \S~\ref{sec:parameters-forbidden}. All energies are in MeV.  For level
    diagrams which illustrate the transitions, see
    Figs.~\ref{fig:levels_MgNaNe} and \ref{fig:levels_NeFO}.}
  \label{tab:transitions}
\end{table}

\section{Analytic Estimates}
\label{sec:analytics}

\citet{Miyaji80} provide a thorough discussion of the different phases
of the evolution of an ONeMg core undergoing compression.  In order to
gain some insight into the relevant physics, we first discuss a simple
model of the evolution up until the onset of thermonuclear oxygen
burning.  In discussing the analytic estimates below, we reference
some of the numerical results from our fiducial MESA model for
comparison.  This model is a cold ONeMg WD ($X_{\mathrm{O}} = 0.5$,
$X_{\mathrm{Ne}} = 0.45$, $X_{\mathrm{Mg}} = 0.05$) accreting at
$\dot{M} = 10^{-6}\,\msunyr$.

\subsection{Overview of evolution}

We have a dense, degenerate core near the Chandrasekhar mass with a
spatially-uniform composition of the carbon-burning products
\oxygen[16], \neon[20] and \magnesium[24], with mass fractions
$X_{\mathrm{O}}$, $X_{\mathrm{Ne}}$, $X_{\mathrm{Mg}}$, respectively.
Fiducially, we choose $X_{\mathrm{O}} = 0.5$,
$X_{\mathrm{Ne}} = 0.45$, $X_{\mathrm{Mg}} = 0.05$.  This is similar
to the central abundances observed in recent calculations of the
evolution of intermediate mass stars that develop these cores
\citep[see e.g., figure 10 of][]{Takahashi13}.  Other recent models of
super-AGB evolution \citep{Farmer15} show typical central magnesium
fractions $X_\mathrm{Mg} \approx 0.03$ in the cases where the carbon
deflagration wave reaches the center (R. Farmer, private
communication).

The degenerate core is ``accreting'' at a rate $\dot{M}$; such
accretion might be set by carbon shell burning in an evolved star,
accretion from a companion in a binary system, or cooling (and the
concomitant reduction in pressure support) of the outer layers of a WD
merger remnant.  The key impact is that the core is being compressed
on a timescale
\begin{equation}
  \label{eq:tcompress}
  t_{\mathrm{compress}} = \left(\frac{d \ln \rho_c}{dt} \right)^{-1} = \left( \frac{d \ln \rho_c}{d \ln M} \right)^{-1} \frac{M}{\dot{M}}~~~.
\end{equation}
For an object supported by degeneracy pressure and in hydrostatic
equilibrium, the central density rises rapidly as one approaches the
Chandrasekhar mass.  Therefore, the compression timescale is
significantly shorter than the timescale for the growth of the core.
For an ideal, zero-temperature white dwarf, in the range
$9 \la \log_{10} \rho_c \la 10$,
\begin{equation}
  \label{eq:tcompress-analytic}
  \frac{d \ln \rho_c}{d \ln M} \approx 28 \left(\frac{\rho_c}{\unit[10^9]{g~cm^{-3}}}\right)^{0.55} ~~~,
\end{equation}
which we obtained by calculating a sequence of models and fitting a
power-law to the results.  This implies
\begin{equation}
  \label{eq:tcompress-numerical}
  t_{\mathrm{compress}} \approx \unit[5\times10^4]{yr}
\left(\frac{\rho_c}{\unit[10^9]{g~cm^{-3}}}\right)^{-0.55} \left(\frac{\dot{M}}{\unit[10^{-6}]{\msun\,\mathrm{yr}^{-1}}} \right)^{-1} ~~~.
\end{equation}
The dynamical time of the white dwarf is extremely short
\begin{equation}
  \label{eq:tdyn}
  t_{\mathrm{dyn}} \approx \frac{1}{\sqrt{G \rho}} \approx 10^{-1} \mathrm{s} \, \left(\frac{\rho}{\unit[10^9]{g\,cm^{-3}}}\right)^{-1/2}
\end{equation}
and so hydrostatic equilibrium will always be preserved (until
collapse ensues, which we do not study in detail in this paper).

\begin{figure}
  \centering
  \includegraphics[width=\columnwidth]{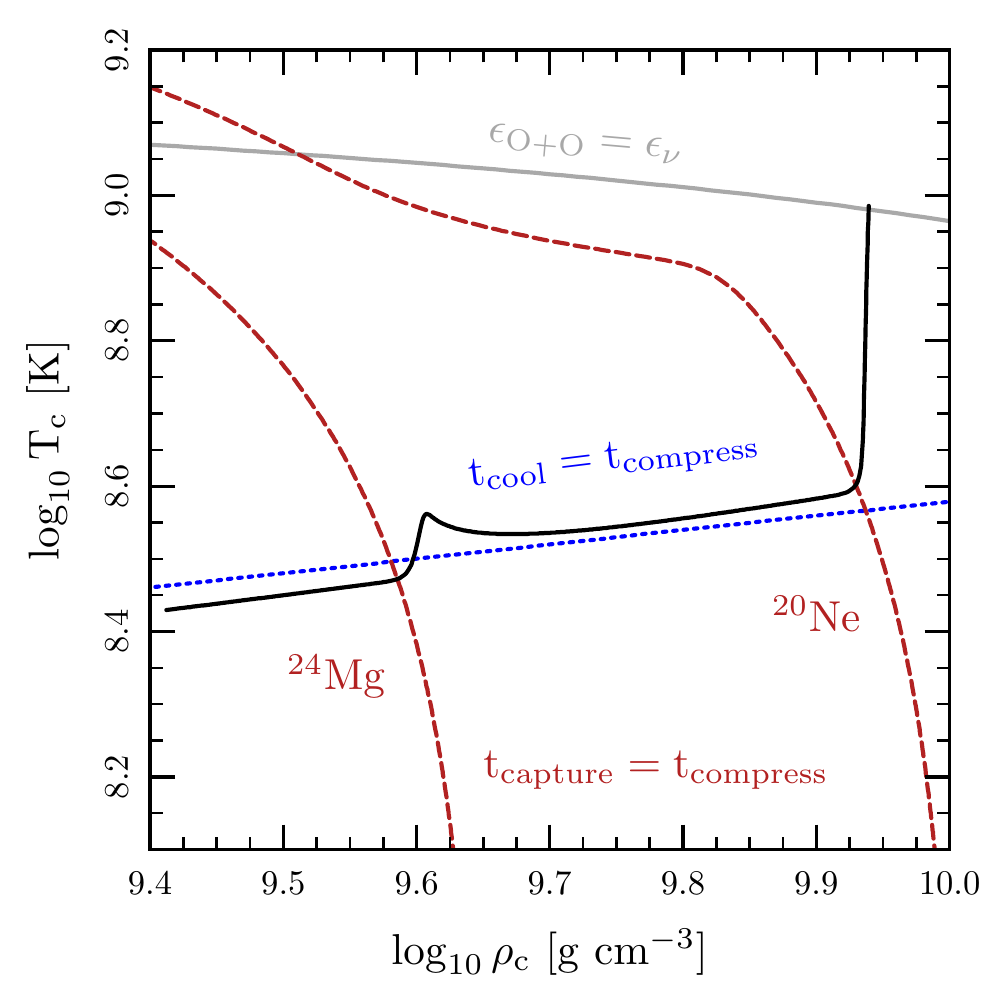}
  \caption{The black solid line shows the central density and
    temperature of the core as it is compressed with a surface
    accretion rate of
    $\dot{M} = \unit[10^{-6}]{\msun}{\mathrm{yr}^{-1}}$ for
    approximately 20000 years of evolution.  The red dashed lines
    indicate when the capture timescales for \magnesium[24] and
    \neon[20] become equal to the fiducial compression time of
    $\unit[10^4]{yr}$.  The blue dotted line shows where the neutrino
    cooling time and compression time are equal.  This balance between
    compressional heating and neutrino cooling determines the thermal
    state of the contracting WD core (aside from brief periods when
    electron captures heat the core). The grey solid line shows where
    the energy generation from thermonuclear oxygen burning exceeds
    the thermal neutrino losses and we stop the calculation.}
  \label{fig:schematic}
\end{figure}

The temperature of the core will be influenced by details of its
previous evolution, such as the accretion history and by the
abundances of isotopes which participate in Urca process cooling.
However, if the compression timescale (and hence overall evolutionary
timescale) is sufficiently slow, heating from compression and cooling
from thermal neutrinos will reach a quasi-equilibrium
\citep{Paczynski71}.  Define the cooling time
\begin{equation}
  \label{eq:cooling}
  t_{\mathrm{cool}} = \frac{c_P T}{\epsilon_\nu},
\end{equation}
where $c_P$ is the specific heat at constant pressure and
$\epsilon_\nu$ is the specific neutrino cooling rate. Then the
relation $t_\mathrm{cool} = t_\mathrm{compress}$ implicitly defines a
temperature for a given density and will characterize the thermal
state of the core aside from periods when e-captures rapidly release
energy.  In Fig.~\ref{fig:schematic}, we show this relation as a blue,
dashed line and demonstrate that our MESA models (the black solid
line) described in \S~\ref{sec:mesa-calculations} and
\S~\ref{sec:parameters} exhibit this relationship.

\subsection{Effects of electron captures}

As the core is compressed, the electron chemical potential increases.
At zero temperature, the electron captures would occur when the Fermi
energy reached the energy difference between the initial and final
nuclear states.  We refer to the density corresponding to this value
of the chemical potential as the threshold density; the terms
sub-threshold and super-threshold reference this density. At non-zero
temperature, even when the chemical potential is below this threshold,
there are some electrons in the high energy tail of the Fermi-Dirac
distribution which are available to capture.  As a result, the
electron capture rate has an exponential dependence on the density and
temperature in the sub-threshold case.

A simple form for the sub-threshold capture rate can be obtained by
expanding equation~\eqref{eq:Iec-fd} in the limit that
$\mu + Q \ll -kT$ (where $\mu$ is the chemical potential and
$Q= Q_g + E_i - E_f$ is the energy difference between the parent and
daughter nuclear state) and assuming that the rate is dominated by a
single transition that begins in the ground state,
\begin{equation}
  \label{eq:lambda-ec-approx}
  \lambda_{\mathrm{ec}} \approx \frac{2 \ln 2}{(ft)}\left(\frac{k T}{m_e c^2}\right)^5 \left(\frac{Q}{k T}\right)^2 \exp\left(\frac{\mu + Q}{kT}\right)~.
\end{equation}

Define the capture timescale to be the inverse of the electron capture
rate $t_{\mathrm{capture}} = \lambda_{\mathrm{ec}}^{-1}$.  The onset
of significant electron captures will occur when the capture time and
the compression time become approximately equal.  Setting
$t_{\mathrm{compress}}$ = $t_{\mathrm{capture}}$ gives an implicit
relationship between $\rho$ and $T$, which is a function of
$\dot{M}$.

At zero temperature, the electron captures would occur at a density
$\rho_\mathrm{ec,0}$ such that
$\mu(\rho_\mathrm{ec,0}) + Q \approx 0$. Solving
equation~\eqref{eq:lambda-ec-approx} for $\mu$ and rewriting the
solution in terms of $\rho$, we find that
$t_\mathrm{compress} = t_\mathrm{capture}$ when
\begin{equation}
\begin{split}
  \label{eq:rho-ec-approx}
  \rho_\mathrm{ec} \approx \rho_{ec,0} &\left[1 + \frac{3 k T}{Q} \times  \right. \\
& \left. \ln \left( 2 \ln 2 \frac{t_\mathrm{compress}}{(ft)}\left(\frac{k T}{m_e c^2}\right)^5 \left(\frac{Q}{k T}\right)^2\right) \right] ~~~,
\end{split}
\end{equation}
where we have neglected the much weaker density dependence of
$t_\mathrm{compress}$ itself.

Equation~\eqref{eq:rho-ec-approx} will be valid up until a temperature
at which the transition rate from an excited state, suppressed by
$\exp(-E_i/kT)$, becomes the dominant contribution to the rate.  As a
rule of thumb, for the transitions we consider, this will happen when
$T \approx E_i/(25 k)$.

\begin{figure}
  \centering
  \includegraphics[width=\columnwidth]{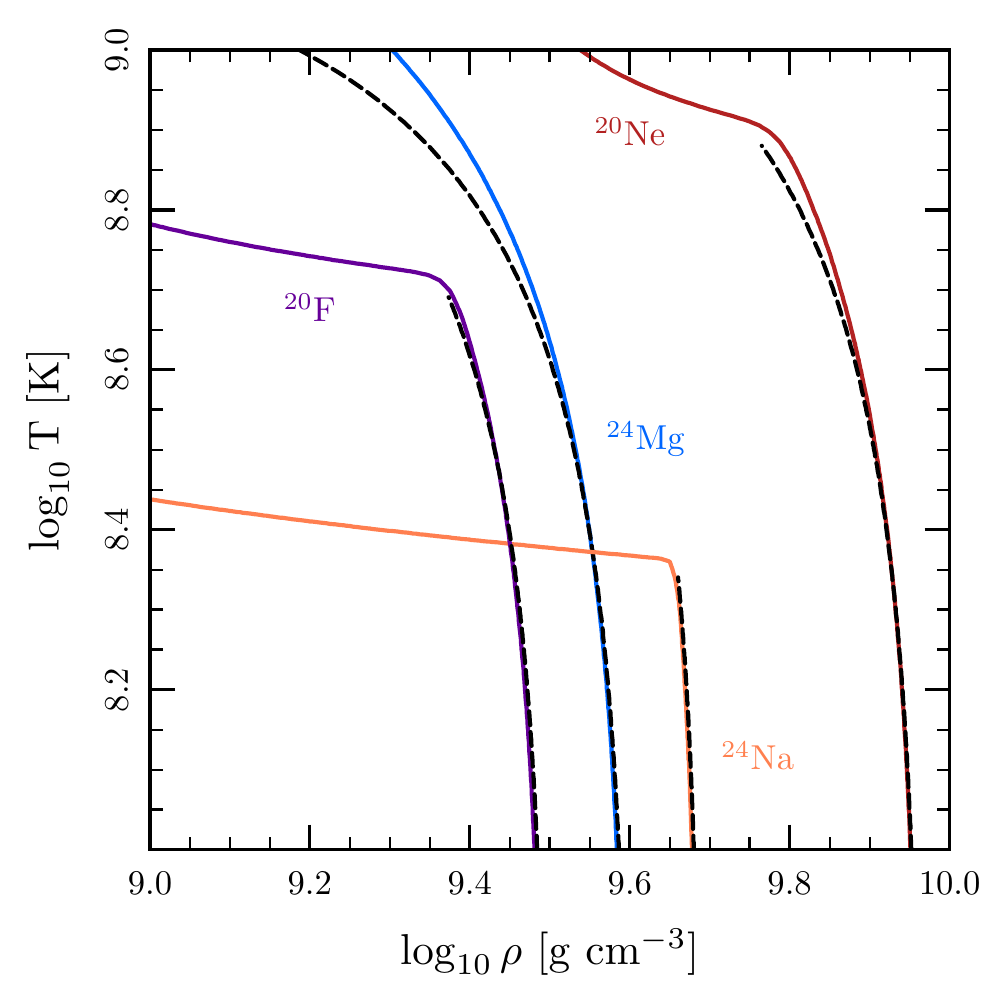}
  \caption{The solid lines show where the electron capture timescale
    is equal to $\unit[10^4]{yr}$ (which is approximately the
    compression timescale of the WD core for
    $\dot{M} = \unit[10^{-6}]{\msun\,\mathrm{yr}^{-1}}$). At $\rho$
    and $T$ greater than those delineated by the solid lines, the
    capture time is less than the compression time. Each line is
    labeled by the name of the isotope undergoing electron capture.
    The black dashed lines show the analytic approximation given in
    equation~\eqref{eq:rho-ec-approx}.  For ease of comparison with
    the analytic results, the Coulomb corrections discussed in
    Appendix~\ref{sec:coulomb} are not present in these calculations.}
  \label{fig:critical-capture}
\end{figure}

Fig.~\ref{fig:critical-capture} shows numerical solutions for the
location in density-temperature space at which
$t_\mathrm{capture} = \unit[10^4]{yr}$ (which is approximately the
compression timescale associated with an
$\dot{M} = \unit[10^{-6}]{\msun\,\mathrm{yr}^{-1}}$) for
\magnesium[24], \sodium[24], \neon[20], and \fluorine[20].  The
approximations for the critical density based on
equation~\eqref{eq:rho-ec-approx} are overlaid as dashed black lines
and are in excellent agreement.  The line for $\fluorine[20]$ is always at
lower density than that of $\neon$, meaning once the first capture in
the $\neon[20] \to \fluorine[20] \to \oxygen[20]$ chain occurs, the
second will immediately follow.  This is not true for $\sodium[24]$
relative to its parent $\magnesium$, meaning the captures in the
$\magnesium[24] \to \sodium[24] \to \neon[24]$ chain will happen at
separate densities when $\log_{10} T \la 8.4$.

The electron captures also influence the temperature evolution of the
core.  When a capture occurs, the chemical potential of the captured
electron, minus the change in nuclear rest mass and the energy in the
emitted neutrinos, is thermalized, heating the plasma.\footnote{A more
  precise definition of the heating rate is given in
  Appendix~\ref{sec:ecapture}, specifically equations \eqref{eq:qec}
  and \eqref{eq:qbeta}.}  This heating is substantial, because the
first capture is often into an excited state (meaning the chemical
potential is higher when the rate of this transition becomes
significant) and the second is typically super-threshold.  Does this
heating drive convection?  If so, this convection will efficiently
transport the entropy out of the core while mixing in fresh fuel for
electron captures.

The electron captures generate entropy, creating a negative radial
entropy gradient in the core.  The captures also reduce the electron
fraction in the core, creating a positive radial gradient in $Y_e$.
The entropy gradient is destabilizing, but the $Y_e$ gradient is
stabilizing.  Simulations which invoked the Schwarzschild criterion
for convection \citep[e.g.][]{Miyaji80}, which does not consider
composition gradients, found that the captures do trigger convection.
Simulations which invoke the Ledoux criterion, which does consider
composition gradients, found that the captures do not trigger
convection \citep[e.g.][]{Miyaji87}.  Hence the different choices lead
to qualitatively different evolution.

The following calculation demonstrates why the electron captures are
unable to trigger convection when accounting for stabilizing
composition gradients.  The Ledoux criterion for convective
instability is
\begin{equation}
  \label{eq:ledoux-criteron}
  \nabla_{\mathrm{ad}} - \nabla_T + B < 0
\end{equation}
where
\begin{equation}
  \label{eq:17}
  B = -\frac{1}{\chi_T}
      \left(\frac{\partial \ln P}{\partial \ln Y_e}\right)_{\rho,T}
      \frac{d \ln Y_e}{d \ln P}~.
\end{equation}
The captures occur over a narrow range in Fermi energy, and hence
density.  Therefore the gradients in $T$ and $Y_e$ across the region
where the captures occur will be large.  This allows us to drop the
$\nabla_\mathrm{ad}$ term.  Replacing the gradients with finite
differences, we then check the inequality
\begin{equation}
  \label{eq:compare-grads}
  \Delta \left(\ln T\right) > -\frac{1}{\chi_T}
      \left(\frac{\partial \ln P}{\partial \ln Y_e}\right)_{\rho,T}
      \Delta\left(\ln Y_e\right).
\end{equation}
For a cold plasma with degenerate electrons and ideal ions,
$\left(\partial \ln P / \partial \ln Y_e\right)_{\rho,T} \approx 4/3$
and $\chi_T \approx 4kT/(\bar{Z} E_\mathrm{F})$.  If a mass fraction
$\Delta X_c$ has undergone electron captures, the associated change in
temperature is
\begin{equation}
  \label{eq:t-change}
  \Delta T \approx \frac{\bar{A}}{A_c}\left(\frac{E_c}{c_P}\right) \Delta X_c~,
\end{equation}
where $A_c$ is the nuclear mass number of the species that is
capturing and $E_c$ is the average energy deposited by a capture.  At
the typical densities and temperatures in our calculation, the ions
are a Coulomb liquid and so $c_P \approx 3k$.  The change in $Y_e$ due
to the captures is
\begin{equation}
  \label{eq:delta-ye}
  \Delta Y_e \approx \frac{\Delta Z}{A_c} \Delta X_c~.
\end{equation}
Both the $A=24$ and $A=20$ chains that we consider are two electron
captures long, so we set $\Delta Z = -2$.

Substituting these estimates into equation~\eqref{eq:compare-grads}
and simplifying, the condition for convective instability becomes
\begin{equation}
  \label{eq:compare-grads-simple}
  \frac{E_c}{E_F} > 2~.
\end{equation}
This inequality demonstrates that in order to trigger convective
instability, the two captured electrons---which each have a
characteristic energy of $E_F$---would have to deposit nearly all
their energy as thermal energy.  This is unrealistic, since
substantial amounts of energy go into the rest mass of the daughter
nucleus and to neutrinos.  From our calculation of the heating rates,
it is clear this inequality is far from being violated: the $A=24$
captures occur at $E_F \approx \unit[6.5]{MeV}$ and release
$E_c \approx \unit[0.5]{MeV}$; the $A=20$ captures occur at
$E_F \approx \unit[8.5]{MeV}$ and and release
$E_c \approx \unit[3]{MeV}$.  Electron captures do not directly
trigger convection.

A region which is Schwarzchild-unstable but Ledoux-stable is
semiconvective.  The semiconvective diffusion coefficient used in MESA
\citep[][following \citealt{Langer83}]{Paxton13} is
\begin{equation}
  \label{eq:semiconvectiveD}
  D_\mathrm{sc} = \alpha_\mathrm{sc}
  \left(\frac{K}{6 c_P \rho}\right)
\left(\frac{\nabla_T-\nabla_\mathrm{ad}}{B + \nabla_\mathrm{ad} -\nabla_T}\right),
\end{equation}
where $K$ is the radiative conductivity.  The values of
$\alpha_\mathrm{sc}$, the semiconvective efficiency adopted in the
literature span the range $10^{-3} \la \alpha_\mathrm{sc} \la 1$
\citep[][citing \citealt{Langer91,Yoon06}]{Paxton13}.

Regions where the electron
captures have not yet occurred and regions where they have completed
do not have a $Y_e$ gradient.  Therefore the width of the
semiconvective zone $H_\mathrm{sc}$ will be roughly the length over
which the density changes by an amount that shifts $E_F$ by $kT$.  We
expect
\begin{equation}
  \label{eq:sc-width}
  H_\mathrm{sc} \sim 4 \left(\frac{kT}{E_F}\right)H_P
\end{equation}
where $H_P$ is the pressure scale height.  Defining
$f = H_\mathrm{sc}/H_P$, we find $f \approx 0.03$ in our MESA models,
consistent with the above estimate.  We define the timescale for
semiconvection to modify the composition and thermal structure in our
models as
\begin{equation}
  \label{eq:semiconvective-timescale}
  t_\mathrm{sc} = \frac{H_\mathrm{sc}^2}{D_\mathrm{sc}} \sim \unit[3 \times 10^{4}]{yr} \left(\frac{1}{\alpha_\mathrm{sc}}\right) \left(\frac{f}{0.03}\right)^2~~~.
\end{equation}
For $\alpha_\mathrm{sc} \la 1$, $t_\mathrm{sc}$ in
equation~\eqref{eq:semiconvective-timescale} is equal to or longer
than time that elapses between $\magnesium$ captures and oxygen
ignitions in our fiducial model.  Moreover, $t_\mathrm{sc}$ is an
upper limit: because of the thinness of the region with a
$Y_e$-gradient, an individual parcel spends less time in a
semiconvective region.  Therefore, we do not consider semiconvection
in our models.

For realistic $\magnesium$ fractions, e-captures on $\magnesium$ do
not release enough energy to initiate thermonuclear fusion.  As a
result, the core continues to compress and we eventually reach the
density where the captures begin on the $A=20$ nuclei.  Once the
capture on \neon[20] occurs, the capture on \fluorine[20] occurs
immediately.  Like the $A=24$ captures, the bulk of the energy
deposition comes from this super-threshold capture, but in this case
the energy per capture is substantially greater,
$E_c \approx \unit[3]{MeV}$.  The characteristic temperature for
oxygen ignition is approximately $\unit[10^9]{K}$ and so from
equation~\eqref{eq:t-change}, we estimate that oxygen will ignite
after an amount $\Delta X_\mathrm{Ne} \approx 0.1$ has undergone
capture.  We halt our main MESA calculations when the energy
generation rate from oxygen burning exceeds the cooling from
neutrinos, implying that a nuclear runaway is assured.

\begin{figure}
  \centering
  \includegraphics[width=\columnwidth]{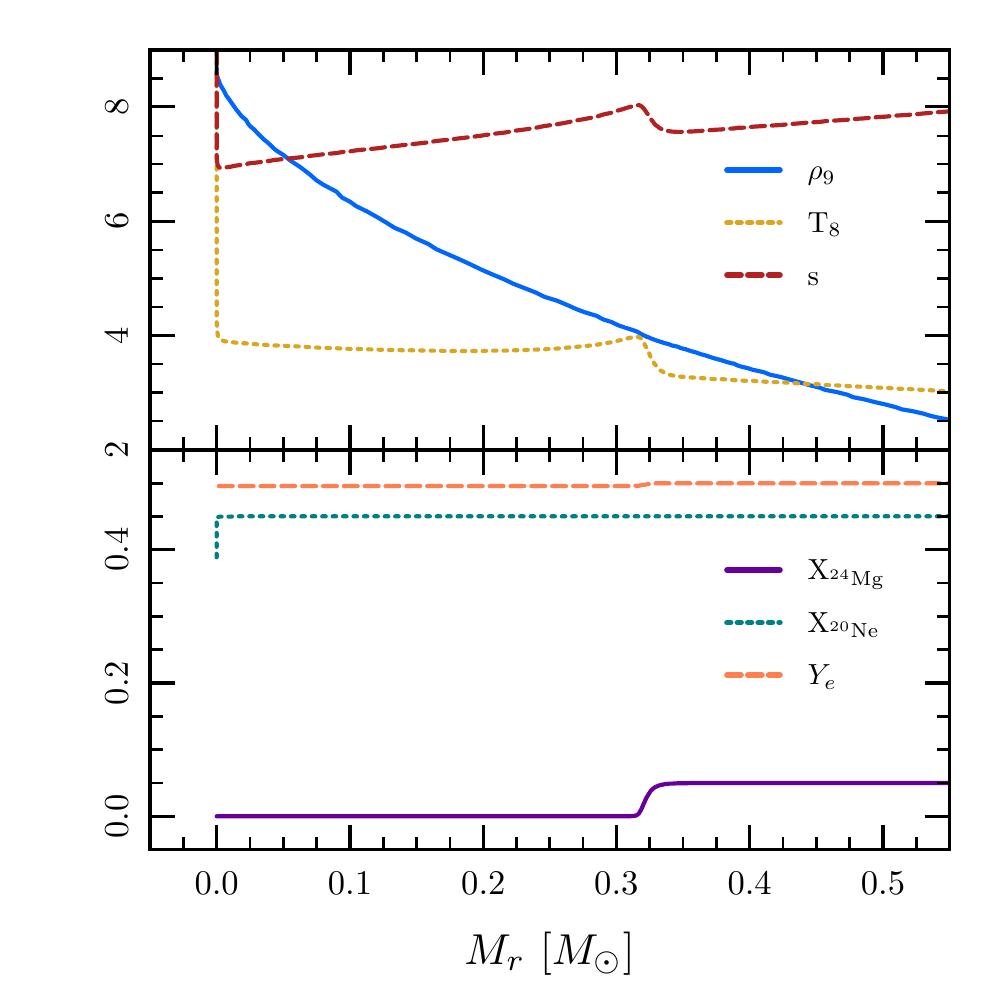}
  \caption{The structure of the model from Fig.~\ref{fig:schematic} at
    the end of the MESA calculation (when the energy generation rate
    from oxygen burning exceeded the neutrino cooling).  The top panel
    shows the density ($\rho_9$, $\rho$ in units of
    $\unit[10^9]{g~cm^{-3}}$), the temperature ($T_8$, $T$ in units of
    $\unit[10^8]{K})$, and the entropy per baryon $s$ (in units of
    $k$), as a function of enclosed mass.  The bottom panel shows
    the mass fractions of $\magnesium[24]$ and $\neon[20]$ as well as
    the electron fraction $Y_e$.  The small region in which there is a
    $Y_e$ gradient due to the $A=24$ captures has been moving outward
    in a Lagrangian sense.  By the time the center reaches the density
    for $\neon$ captures, the inner $0.3 \msun$ has already been
    depleted of $\magnesium[24]$ due to e-captures.The subsequent
    evolution is discussed in \S~\ref{sec:discussion}.}
  \label{fig:final-profile}
\end{figure}

We have focused primarily on the evolution of the center of the core,
but the density of the rest of the core increases during
compression. Electron captures on the $A=24$ elements have been
occurring off-center as parcels of the star reach conditions favorable
for these captures.  This is illustrated in
Fig.~\ref{fig:final-profile}, where one can see the depletion of
$\magnesium$ in the inner $0.3 \msun$ of the star.

\subsection{Thermal runaway from $^{20}$Ne Captures and Oxygen
Deflagration Initiation}

\begin{figure}
  \centering
  \includegraphics[width=\columnwidth]{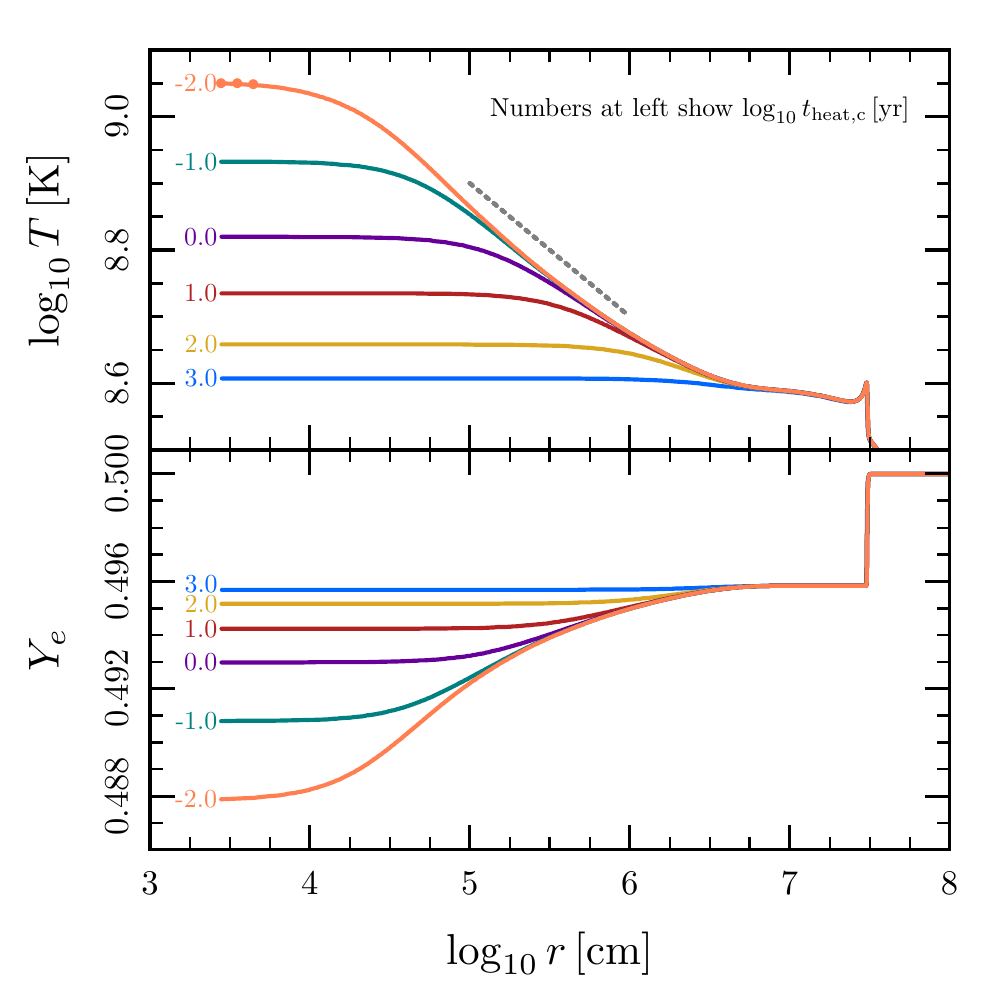}
  \caption{The temperature ($T$) and electron fraction ($Y_e$)
    profiles in the MESA model shown in Fig.~\ref{fig:final-profile},
    as it is approaching the onset of O fusion. At these temperatures,
    the energy release is dominated by the $A=20$ captures, so $T$ and
    $Y_e$ are closely tied.  The lines are labeled by the heating
    timescale at the center of the model.  The dotted grey line in the
    top panel shows the slope of the temperature profile expected for
    a thermal runaway with diffusion.  The mass resolution in this
    calculation is significantly higher than that in other figures in
    order to resolve the small region (regulated by thermal diffusion)
    within which thermal runaway sets in.  The dots on the orange
    (hottest) temperature curve indicate the locations of the
    innermost three MESA zones; the mass in the central zone is
    roughly $4 \times 10^{-13} \msun$.}
  \label{fig:gradients}
\end{figure}

In Fig.~\ref{fig:gradients} we show the evolution of the center of our
MESA models as the $A=20$ captures begin.\footnote{The MESA run shown
  in this plot used a finer central spatial and temporal resolution
  than our fiducial case in order to better resolve the onset of these
  steep central gradients.}  The profiles are labeled by the central
heating time of the model,
$t_{\rm heat,c} = c_pT / \epsilon_\mathrm{nuc}$.  At these
temperatures, the energy generation rate is dominated by the $A=20$
captures, which are undergoing a thermal runaway in the thermally
conducting core. From equation~\eqref{eq:t-change}, the
change in $Y_e$ associated with increasing the temperature from its
value before the $A=20$ captures, $T \approx \unit[4\times10^{8}]{K}$,
to the temperature for oxygen ignition, $T \approx \unit[10^{9}]{K}$,
is $\Delta Y_e \approx 0.006$, in good agreement with the
change observed in the lower panel of Fig.~\ref{fig:gradients}.
Changes in $T$ and $Y_e$ will no longer be so tightly coupled once
energy release from oxygen fusion exceeds that from electron
captures, pushing the core towards convective instability. However, in
order to reach convective instability, equation~\eqref{eq:compare-grads}
requires
\begin{equation}
  \Delta T > \frac{\bar{A} E_F}{3 k} \Delta Y_e \approx
  \unit[3.5\times10^{9}]{K} \left(\frac{\Delta Y_e}{0.006}\right)~,
\end{equation}
a temperature so large that the central heating timescale from oxygen
fusion would be $t_\mathrm{heat,c} \approx \unit[10^{-5}]{s}$.  We
show here that a thermal runway is triggered long before such a
condition is reached.

The $A=20$ electron captures occur in an environment where the
electron Fermi energy is below the energy threshold. In this
sub-threshold case, those electrons that capture are on the thermal
tail of the distribution, making the rate very sensitive to both
density and temperature.\footnote{Because the bulk of the heating
  comes from the super-threshold electron capture on $\fluorine[20]$
  that immediately follows the $\neon[20]$ capture, the capture rate
  on $\neon[20]$ is a good proxy for the temperature and density
  dependence of the heating rate.} We now show that this naturally
leads to a local thermal runaway in the core whose size is limited by
thermal conduction. This runaway provides the ``hot-spot'' needed to
initiate the oxygen deflagration from the center of the star.

The strong density sensitivity of the $A=20$ captures
implies that the runaway will begin at the exact center of the isothermal
core. However, the pressure declines away from the core, leading to
a temperature gradient on the
scale over which the electron capture rate (and hence the heating
rate) varies by order unity. In this  sub-threshold case,
$d\ln\lambda/d\ln P = E_F / (4kT)$, so the change in pressure needed
to have the rate be less at the outer edge than the center is
\begin{equation}
  \label{eq:p-change}
  \Delta P \approx 4 P_c \left(\frac{kT}{E_F}\right) \approx
  \left(\frac{\rho_c Y_e}{m_p}\right) k T_c~.
\end{equation}
Hydrostatic equilibrium provides such a pressure change
over a length scale
\begin{equation}
  \label{eq:core-size}
  l_T = \left(\frac{3 \Delta P}{2 \pi G \rho_c^2}\right)^{1/2} \approx
  \unit[4\times10^{6}]{cm}~,
\end{equation}
where we have used $T_c \approx \unit[4 \times 10^8]{K}$ and
$\rho_c \approx \unit[9 \times 10^{9}]{g\,cm^{-3}}$, corresponding to
the onset of $A=20$ captures in our fiducial model.  This estimate is
consistent with the length scale observed at the onset of the
runaway in Fig.~\ref{fig:gradients}.

The subsequent evolution of the runaway is driven by the temperature
sensitivity of the sub-threshold electron capture rate. This rate is
well-approximated by equation~\eqref{eq:lambda-ec-approx}, yielding a
logarithmic derivative of the rate with respect to temperature of
\begin{equation}
  \label{eq:3}
  \frac{d \ln \lambda}{d \ln T} = 3 - \frac{\mu + Q}{kT} ~.
\end{equation}
Physically, the second term is the how far the transition is below its
threshold energy (in units of $kT$).  As in the $A=24$ case,
electron captures become important when
$\lambda_\mathrm{ec}^{-1} \approx t_\mathrm{compress}$; from
equation~\eqref{eq:lambda-ec-approx} this occurs at
$(\mu + Q)/(kT) \approx -14$. The thermal runaway is
sufficiently rapid that $\mu + Q$ remains approximately fixed.  This
implies that the captures will be extremely temperature sensitive,
scaling as $\lambda \propto T^n$, where
\begin{equation}
  \label{eq:6}
  n \equiv \frac{d \ln \lambda}{d \ln T} \approx 3 + 14
  \left(\frac{T}{\unit[4\times10^8]{K}}\right)^{-1}~.
\end{equation}
In the following estimates, we will take $n \approx 12$, and since
$n \gg 1$, we will treat $n \approx n \pm 1$.

Hence, as captures begin, their density dependence leads to a temperature
gradient on the length scale given by equation~\eqref{eq:core-size}.
Because convection is not initiated, the heating remains local, and
this temperature gradient will grow with time in a thermal runaway.
Once it is sufficient
to cause an order unity variation of the capture rate across a given
length $r$, the gradient will become non-linear.  The hotter part will
begin to evolve more rapidly and the evolution of the cooler part will
freeze-out.  This will occur when $dT/dr \approx (T/n)/r$ and so on
its own, thermal runaway leads to a characteristic profile where
$d\ln T/d\ln r \approx 1/n$. However, thermal
conduction limits the volume that can runaway to a fixed temperature,
keeping regions where $t_\mathrm{th} \la t_\mathrm{heat}$ approximately
isothermal. The thermal diffusivity from electron conduction is
$D_\mathrm{th} \approx \unit[30]{cm^2\,s^{-1}} (T / \unit[10^9]{K})$,
meaning that the timescale for conduction to modify the thermal
structure over a lengthscale $r$ is
\begin{equation}
  \label{eq:thermal-timescale}
  t_\mathrm{th} = \frac{r^2}{D_\mathrm{th}} \approx \unit[10^{3}]{yr}
  \left(\frac{r}{\unit[10^6]{cm}}\right)^2
  \left(\frac{T}{\unit[10^{9}]{K}}\right)^{-1}~.
\end{equation}
Therefore, the size of the isothermal region at the center
of the model scales like $r \propto T_c^{1-n/2}$.  Thus, as the
runaway
progresses, it will create a temperature profile with
$d\ln T/d\ln r \approx -1/5$.  The dotted grey line in the top panel
of Fig.~\ref{fig:gradients} shows this slope, which agrees well with
the temperature evolution in the MESA calculations.
The semiconvective instability grows on the thermal diffusion time.
During the thermal runaway, by definition,
$t_\mathrm{heat} \la t_\mathrm{th}$. Hence, the evolution of
the core during this phase will be sufficiently fast that
semiconvection will not modify the temperature or composition.

This thermal runaway leads to a small volume at the core reaching very
high tempertures, eventually to values large enough for heating from
oxygen fusion to play a role. This occurs when
$T \approx \unit[1.1\times10^{9}]{K}$, where the heating time is
$t_\mathrm{heat} \approx \unit[10^{-2}]{yr}$. From
equation~\eqref{eq:thermal-timescale}, the hottest (isothermal, so
$t_\mathrm{th} \approx t_\mathrm{heat}$) part of the core will have a
size $r \approx \unit[3\times 10^{3}]{cm}$, which encompasses about
$3 \times 10^{-13}$ of the total mass.  The finest central zoning that
we were able to achieve in our MESA calculations (as shown in
Fig.~\ref{fig:gradients}) was a mass resolution of approximately
$4 \times 10^{-13} \msun$.  Therefore, just as oxygen burning begins
to dominate the energy release, the small size of this region prevents
us from continuing to follow its evolution in our full star MESA
simulations.

The conditions created in the core of the star as energy generation by
oxygen fusion begins to dominate over $A = 20$ captures lead naturally
to the development of an oxygen deflagration wave.  In particular, we
have shown that oxygen fusion begins in a region at the core of the
star whose size is determined by $t_\mathrm{th} \la t_\mathrm{heat}.$
With $t_\mathrm{heat}$ identified as the heating time associated with
oxygen fusion, this is precisely the condition for the onset of an
oxygen deflagration wave; \citet{Timmes92} defined the deflagration
``trigger mass'' to be the mass contained within the region satisfying
this constraint.  Therefore, we are confident that the hot central
region present at the end of our MESA calculations, being unstable to
thermal runaway, will continue to grow in temperature and shrink in
size, eventually reaching the laminar deflagration solutions of
\citet{Timmes92}.\footnote{At the density in our MESA calculations, the
laminar deflagration width is
$\delta \approx \unit[3\times10^{-5}]{cm}$, far below our ability to
resolve in our full star simulations.}
The outgoing deflagration wave will sweep across
this thermally unstable core in less than one second.

It is important to stress that the onset of the oxygen deflagration in
the AIC context is substantially different than the ``simmering
phase'' in single degenerate Type Ia supernovae progenitors.  There,
after pycnonuclear carbon ignition occurs, the entropy release from
carbon burning drives the formation of a central convection zone.  The
growth and heating of this convective zone lead to a significant
decrease in the central density between the time of carbon ignition
and the development of a deflagration.

In our models, by contrast, no central mixing occurs because of the
stabilizing effect of the composition gradient associated with $A=20$
captures. Therefore the central density at which oxygen ignition
occurs, and at which we halt our MESA calculations, is a good estimate
of the central density at which the oxygen deflagration develops.  We
discuss the propagation of this deflagration and its influence on the
final outcome in \S~\ref{sec:discussion}.

\section{Details of MESA Calculations}
\label{sec:mesa-calculations}

All of the calculations performed in this paper are based on revision
6596 (released 2014-06-07), with some modifications to support our
weak rate calculations.  The incorporation of these changes into the
mainline MESA code will be discussed in the upcoming MESA Instrument
Paper III (Paxton et al.~2015). As required by the MESA manifesto, the
inlists and source code modifications necessary to reproduce our
calculations will be posted on http://mesastar.org.

\subsection{Generation of Initial Models}

In order to perform the parameter study discussed
\S~\ref{sec:parameters}, it is necessary to have a set of models of
ONeMg cores with a range of different temperatures and compositions to
use as initial conditions.  We generate an idealized set of models via
the following \textit{ad hoc} steps.  During each step, all nuclear
reactions are turned off, ensuring that the model will continually
contract until halted by degeneracy pressure.

We begin with a $1.325 \msun$ pre-main sequence model of normal
(roughly solar) composition.  We evolve this model until it reaches a
central density of $\log_{10} \rho = 3$ (cgs).  We then relax the
(homogeneous) composition to the desired $\oxygen$, $\neon$, and
$\magnesium$ mass fractions and allow the model to evolve until the
central density reaches $\log_{10} \rho = 7$.  Then we set the model
to accrete at a constant $\dot{M}$ and evolve until the central
density reaches $\log_{10} \rho = 9.4$, which is still below the
threshold for the onset of the electron capture reactions of interest.
In order to achieve different core temperatures, we vary $\dot{M}$;
models with higher (lower) accretion rates have less (more) time for
neutrino cooling to carry away energy and are correspondingly hotter
(colder).  By this means, we arrive a set of models with varied
compositions and central temperatures to use as initial models.

\subsection{Important MESA Options}

While our full inlists will be made publicly available, we highlight
some of the most important MESA options used in the calculations.
This section assumes the reader is familiar with specific MESA
options.  Please consult the instrument papers \citep{Paxton11,
  Paxton13} and the MESA website\footnote{http://mesa.sourceforge.net}
for a full explanation of the meaning of these options.

Since MESA is an implicit code, it is important that we choose
timesteps that will resolve the processes of interest.  The evolution
of the ONeMg cores is driven by the increase in central density (and
hence Fermi energy) caused by the ongoing compression.  Therefore, our
default runs include a timestep criterion based specifically on
changes in central density
\begin{verbatim}
    delta_lgRho_cntr_hard_limit = 3e-3
    delta_lgRho_cntr_limit = 1e-3
\end{verbatim}
in addition to the primary spatial and temporal convergence settings
of
\begin{verbatim}
    varcontrol_target = 1e-3
    mesh_delta_coeff = 1.0 .
\end{verbatim}
Evidence demonstrating that this set of MESA options yields a
converged result is shown in Appendix~\ref{sec:convergence}.

These calculations use a nuclear network based on the
\texttt{co\_burn.net} network included with MESA with the addition of
the isotopes $\oxygen[20]$, $\fluorine[20]$, $\neon[24]$, and
$\sodium[24]$ and the weak reactions linking the $A=20$ isotopes to
$\neon[20]$ and the $A=24$ isotopes to $\magnesium[24]$.  The special
treatment of these weak reactions (as discussed in
Appendix~\ref{sec:ecapture}) is activated with the options
\begin{verbatim}
    use_special_weak_rates = .true.
    ion_coulomb_corrections = 'PCR2009'
    electron_coulomb_corrections = 'Itoh2002'
\end{verbatim}
where the last two lines select the Coulomb corrections discussed in
Appendix~\ref{sec:coulomb}.

\section{Parameter Studies}
\label{sec:parameters}

In this section, we use a suite of MESA calculations to study how a
variety of parameters affect the evolution of these cores. The key
question we will answer is whether reasonable variation in these
parameters will affect the final outcome.

The first parameter (\S~\ref{sec:parameters-xmg}), the initial
$\magnesium$ mass fraction ($X_\mathrm{Mg}$), is an intrinsic property
of the ONeMg core, set during the process that produced the core.
Variation in this value may reflect variation in the formation process
(e.g., the initial mass of the star that produced it) as well as
limits of our knowledge (e.g., uncertainties in quantities such as the
$\carbon(\alpha,\gamma)\oxygen$ reaction rate).  The second parameter
(\S~\ref{sec:parameters-tc-mdot}), the accretion rate $\dot{M}$, is
set by the current state of the system (e.g., the properties of a
binary companion, the details of shell-burning).  The third parameter
(\S~\ref{sec:parameters-forbidden}), the strength of the second
forbidden transition between the ground states of $\neon$ and
$\fluorine[20]$, reflects a limit in our current knowledge.

\subsection{Effect of a \magnesium[24] mass fraction}
\label{sec:parameters-xmg}

\begin{figure}
  \includegraphics[width=\columnwidth]{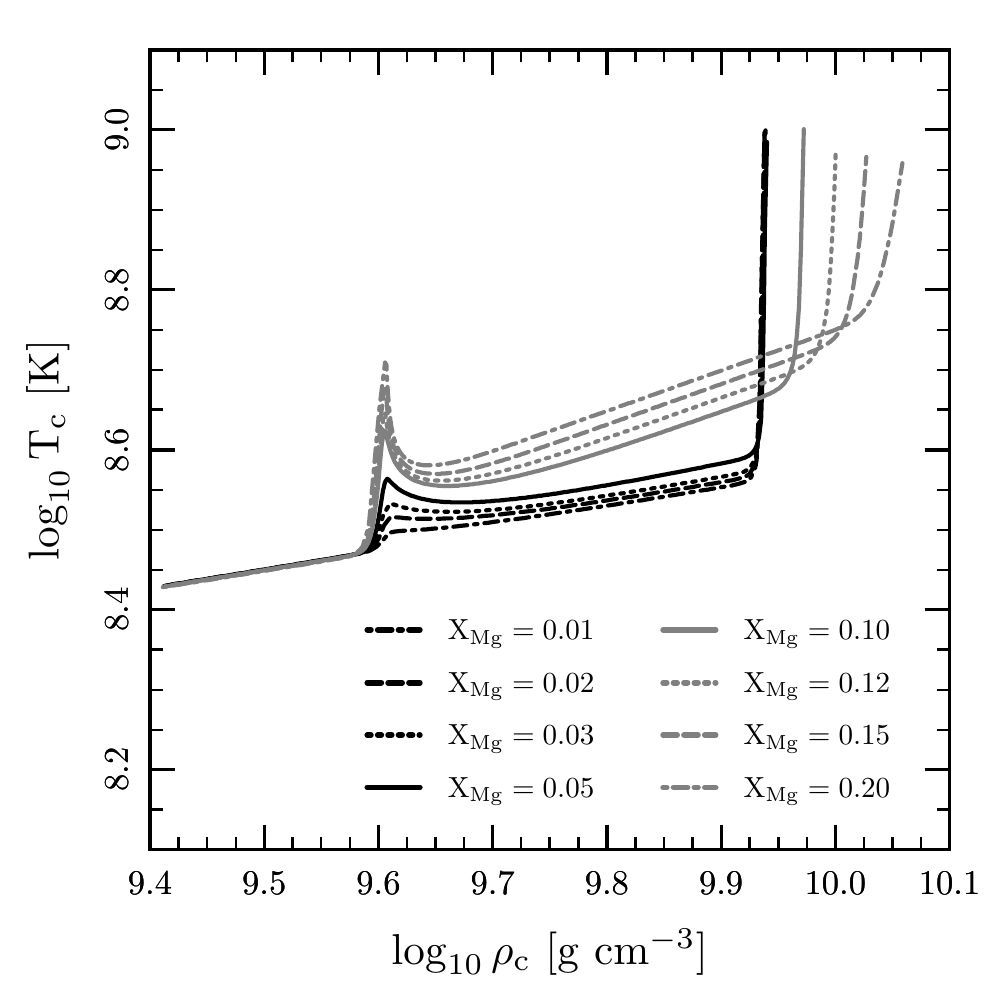}
  \caption{The evolution of the central density and temperature for
    different magnesium mass fractions.  At lower $X_\mathrm{Mg}$
    values, the density at which oxygen ignition occurs (the end of
    the track) is independent of $X_\mathrm{Mg}$; at higher
    $X_\mathrm{Mg}$ values, the density at which oxygen ignition
    occurs increases with increasing $X_\mathrm{Mg}$.  The text and
    Figs.~\ref{fig:compression}~and~\ref{fig:critical_xmg} explain
    the origin of this trend.}
  \label{fig:xmg}
\end{figure}

\citet{Gutierrez05} performed a parameter study of the effects of the
\magnesium[24] mass fraction. We follow their approach of varying the
central \magnesium[24] fraction, while holding the
\oxygen[16]/\neon[20] ratio fixed.\footnote{In accordance with our
  fiducial model, we set this ratio be at $10/9$.  The choice of this
  ratio does slightly influence the density at which captures occur
  through the dependence of the Coulomb corrections on $\bar{Z}$.
  However, as a small shift on top of a small shift, we do not explore
  variations in this ratio.}  We explore a wide range of
\magnesium[24] mass fractions, from 0.01 up to 0.20.  This latter
value is well above the expected \magnesium[24] fraction given current
reaction rates.  Our results are shown in Fig.~\ref{fig:xmg}.

The temperature increase due to the $A = 24$ electron captures scales
roughly linearly with $X_\mathrm{Mg}$, as expected from
equation~\eqref{eq:t-change}.  For $X_\mathrm{Mg} \la 0.07$, neutrino
cooling erases the effect of the heating and the trajectories converge
back towards the $t_\mathrm{cool} = t_\mathrm{compress}$ relation
described in \S~\ref{sec:analytics}.  Correspondingly, the density at
which the captures on $\neon$ occur---and thus the density at which
oxygen ignites---is independent of $X_\mathrm{Mg}$.  For
$X_\mathrm{Mg} \ga 0.07$, an increase in $X_\mathrm{Mg}$ leads to the
onset of $\neon$ captures (and oxygen ignition) at a higher
density. At even higher values (not shown), the heating from the
$A=24$ captures is sufficient to directly ignite oxygen burning as
noted by \citet{Miyaji87} and \citet{Gutierrez05}.

In the limited set of models presented in \citet{Gutierrez05} this
bifurcation in the core evolution around $X_\mathrm{Mg} \approx 0.07$
is not evident.  However, it has a clear physical explanation.  In
Appendix~\ref{sec:ztwd} we discuss a simple model of a
zero-temperature white dwarf with a low-$Y_e$ core that explains these
results; here, we demonstrate the consistency of these calculations
with our MESA models, eliding the details.

\begin{figure}
  \centering
  \includegraphics[width=\columnwidth]{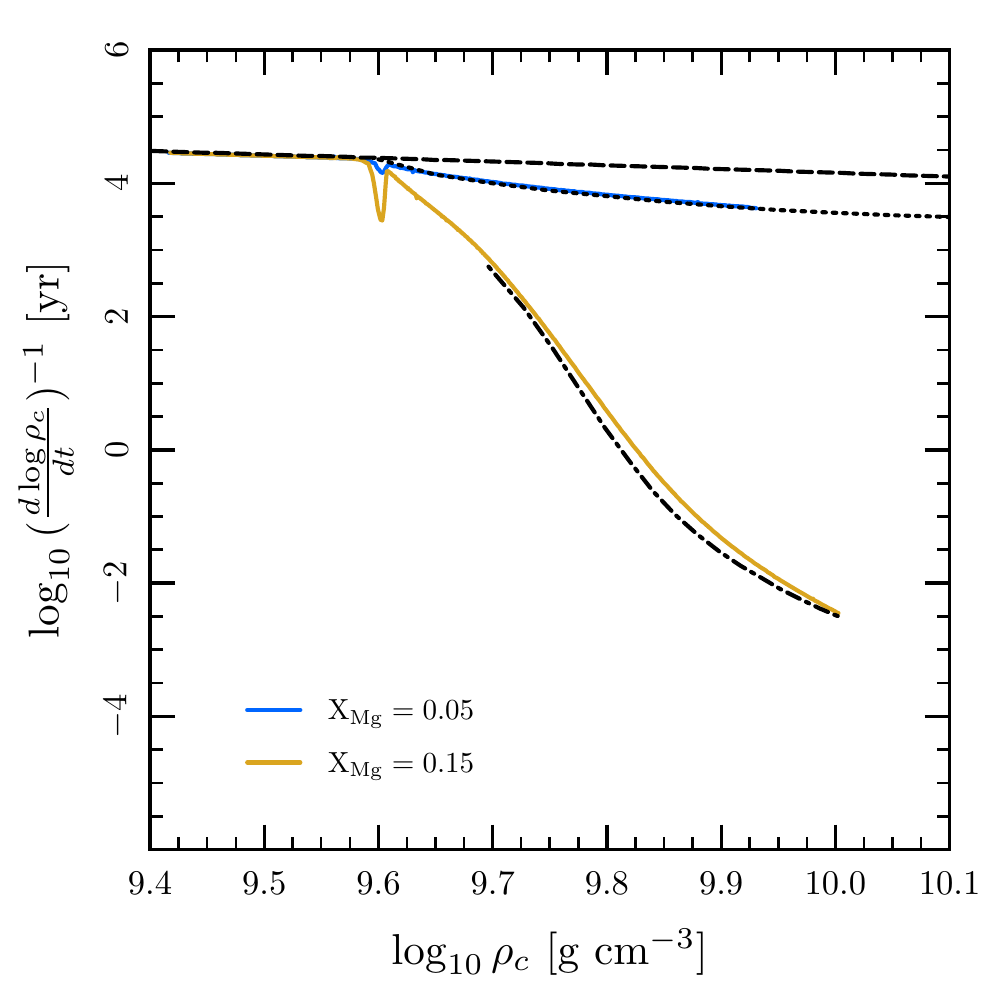}
  \caption{The compression time of selected MESA models.  The model
    with $X_\mathrm{Mg} = 0.05$ is compressed at a rate controlled by
    the accretion rate. The dashed line shows the compression rate
    given in equation~\eqref{eq:tcompress-numerical} and the dotted
    line shows the compression rate expected for a zero temperature
    white dwarf in which $Y_e$ suddenly falls at
    $\log_{10}\rho = 9.6$, due to the electron captures on
    $\magnesium$ and $\sodium[24]$.  The agreement demonstrates that
    the central density is controlled by the total mass.  The model
    with $X_\mathrm{Mg} = 0.15$ experiences a much more dramatic
    decrease in the compression time because of the larger decrease in
    $Y_e$ (see Fig.~\ref{fig:critical_xmg}).  The black dash dotted
    line shows the neutronization timescale expected from the
    calculations in Appendix~\ref{sec:ztwd}, where the central density
    is evolving at fixed mass.  Note that in both cases the
    compression timescale still remains orders of magnitude longer
    than the dynamical time.}
  \label{fig:compression}
\end{figure}

Fig.~\ref{fig:compression} shows the compression time in two models,
one on each side of this threshold value of $X_\mathrm{Mg}$.  The core
evolution for both models follows the dashed black line defined by
equation~\eqref{eq:tcompress-numerical} up until the onset of the
$A=24$ electron captures at $\log_{10} \rho_c \approx 9.6$.  Above
this density, the $X_\mathrm{Mg} = 0.05$ model (blue line) follows the
dotted line, which is the track expected for the fiducial value of
$\dot{M}$ and the value of $d\ln M/d\ln \rho_c$ in
equation~\eqref{eq:tcompress} calculated from a zero-temperature white
dwarf model in which $Y_e$ decreases for $\log_{10} \rho > 9.6$.  See
Fig.~\ref{fig:ztwd2-xmg-0p05} and surrounding discussion for the
details of this zero-temperature model.

In the $X_\mathrm{Mg} = 0.15$ model (yellow line in
Fig.~\ref{fig:compression}), once the $\magnesium$ captures occur at
$\log_{10} \rho_c \approx 9.6$, the compression timescale begins to
fall dramatically.  By the time $\neon$ capture densities are reached
($\log_{10} \rho \approx 10$), the compression timescale is orders of
magnitude smaller than in the lower $X_\mathrm{Mg}$ models, though it
remains significantly longer than the dynamical time.  Recall that
significant electron captures only occur when the capture time
satisfies the relation $t_\mathrm{compress} = t_\mathrm{capture}$;
this means that for shorter compression timescales, the core must
reach higher densities, and hence higher capture rates, before the
effects of the captures become apparent.  As $X_\mathrm{Mg}$
increases, models experience a larger drop in $Y_e$, and compress more
quickly.  This explains the trend of increasing oxygen ignition
density with increasing $X_\mathrm{Mg}$ seen in Fig.~\ref{fig:xmg}.

\begin{figure}
  \centering
  \includegraphics[width=\columnwidth]{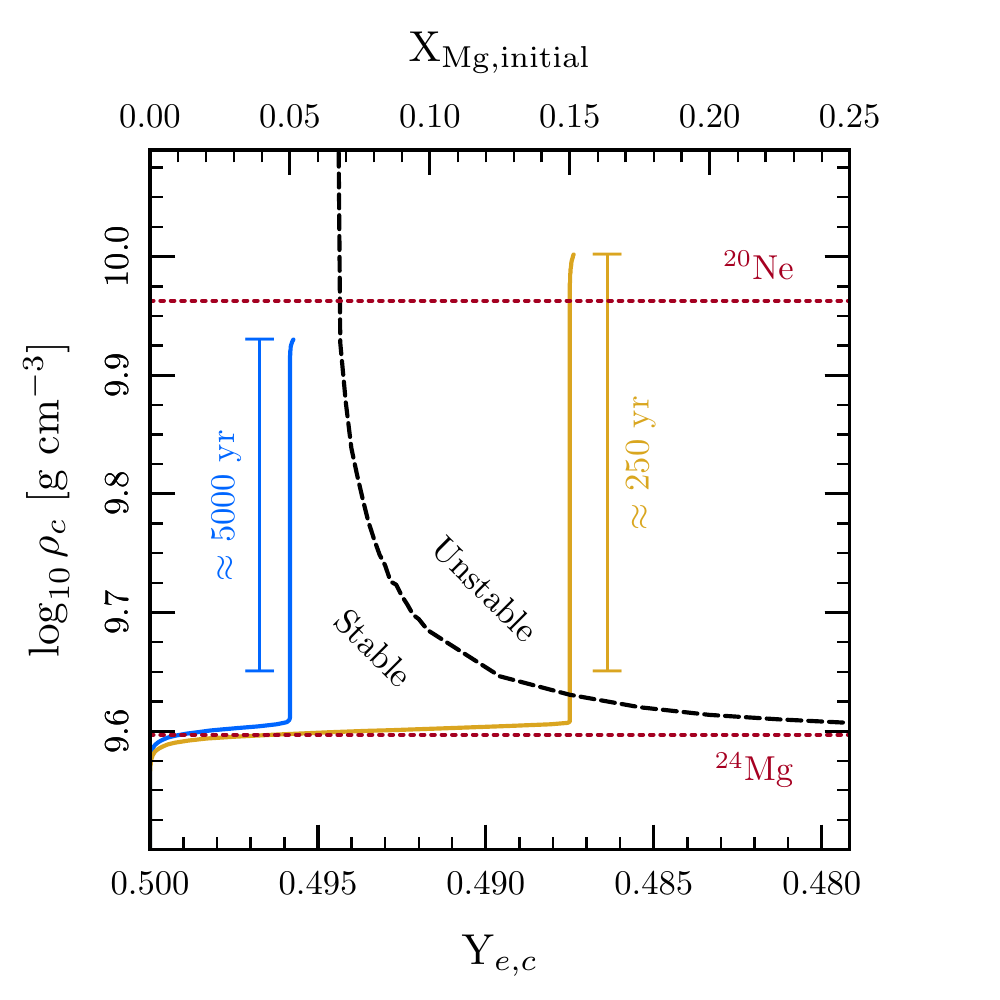}
  \caption{The evolution of the central electron fraction and central
    density for two values of $X_\mathrm{Mg}$ (blue and yellow solid
    lines).  The time elapsed during the evolution from
    $\log_{10}\rho = 9.65$ to the onset of $\neon$ captures is
    indicated next to each track.  The dashed line shows the stability
    curve for a zero temperature white dwarf which neutronizes to the
    value of $Y_{e,c}$ shown on the x-axis at a density of
    $\log_{10}\rho = 9.6$. The dotted red lines show the threshold
    electron capture densities for $\neon$ and $\magnesium$.  For
    $X_\mathrm{Mg} \ga 0.07$, the onset of $A=24$ electron captures
    reduces $Y_e$ such that subsequent compression drives the
    equivalent zero temperature models into a dynamically unstable
    region of parameter space.  Past this point, the contraction
    accelerates significantly (as shown in
    Fig.~\ref{fig:compression}).  In these cases, $\magnesium$
    captures alone have assured collapse.}
  \label{fig:critical_xmg}
\end{figure}

To physically understand the different evolution of the
$X_\mathrm{Mg} \ga 0.07$ models, we consider an idealized model of the
effect of Mg captures on the structure and stability of the ONeMg
core.  We assume that the $A=24$ electron captures occur
instantaneously above a density of $\rho_n = 9.6$.  Using the approach
described in Appendix~\ref{sec:ztwd}, we can then determine the
central density of the zero temperature model with the maximum mass.
The result of this calculation is shown as the dashed line in
Fig.~\ref{fig:critical_xmg}. Any model with $X_\mathrm{Mg} \ga 0.07$
will cross the stability line before the onset of $\neon$ captures.
Moreover, for a larger change in $Y_e$ (associated with a larger
$X_\mathrm{Mg}$ in the current example), the onset of instability
occurs at lower central density.  These results explain the
qualitatively different behavior of the high $X_\mathrm{Mg}$ models in
Fig.~\ref{fig:xmg}.

Above the dashed line in Fig.~\ref{fig:critical_xmg}, the zero
temperature models are dynamically unstable and would contract on the
dynamical timescale.  But the characteristic electron capture
timescales are longer than the dynamical time, and so the assumption
that the captures are effectively instantaneous (used in the idealized
models in Appendix D) does not hold in the real MESA models.  As the
contraction timescale gets shorter, only material at densities where
the capture timescale is shorter than the contraction timescale can
have had significant captures.  Therefore the density above which the
captures have completed, $\rho_n$, shifts to higher values.  There is
no longer time for the total mass to change and so the timescale for
the evolution of the central density is no longer set by the accretion
rate.  Instead, the core compresses on the significantly shorter
neutronization timescale,
\begin{equation}
  \label{eq:tneut-mg}
  t_n = \left(\frac{d \ln Y_e}{dt}\right)^{-1} = \frac{Y_e A_\mathrm{Mg}}{X_\mathrm{Mg}\lambda_\mathrm{ec}} \approx 80 \left(\frac{X_\mathrm{Mg}}{0.15}\right)^{-1}t_\mathrm{capture}.
\end{equation}
Given a fixed $M$, the models in Appendix~\ref{sec:ztwd} give a
relationship between $\rho_c$ and $\rho_n$.  Calculating $t_n$ by
evaluating $t_\mathrm{capture}$ at the $\rho_n$ corresponding to each
$\rho_c$ gives the dash dotted line in Fig.~\ref{fig:compression},
which agrees well with the result of the MESA calculation.  See
Fig.~\ref{fig:ztwd2-xmg-0p15} and surrounding discussion for the
details of these zero-temperature models.

\subsection{Effect of central temperature and accretion rate}
\label{sec:parameters-tc-mdot}

As illustrated in Fig.~\ref{fig:critical-capture}, the density at
which electron captures begin is temperature dependent.  Our fiducial
model begins at a central density $\log_{10}\rho_c \approx 9.4$ and
$\log_{10} T_c \approx 8.4$.  This central temperature is a free
parameter, but as discussed in \S~\ref{sec:analytics}, a new central
temperature will be established by the balance between neutrino
cooling and compressional heating. Therefore, the central temperature
when captures occur (in particular the $A=20$ captures, and quickly
thereafter oxygen ignition) is weakly dependent on the initial
temperature.  This fact makes it difficult to separately illustrate
the effects of the temperature and accretion rate.

\begin{figure}
  \centering
  \includegraphics[width=\columnwidth]{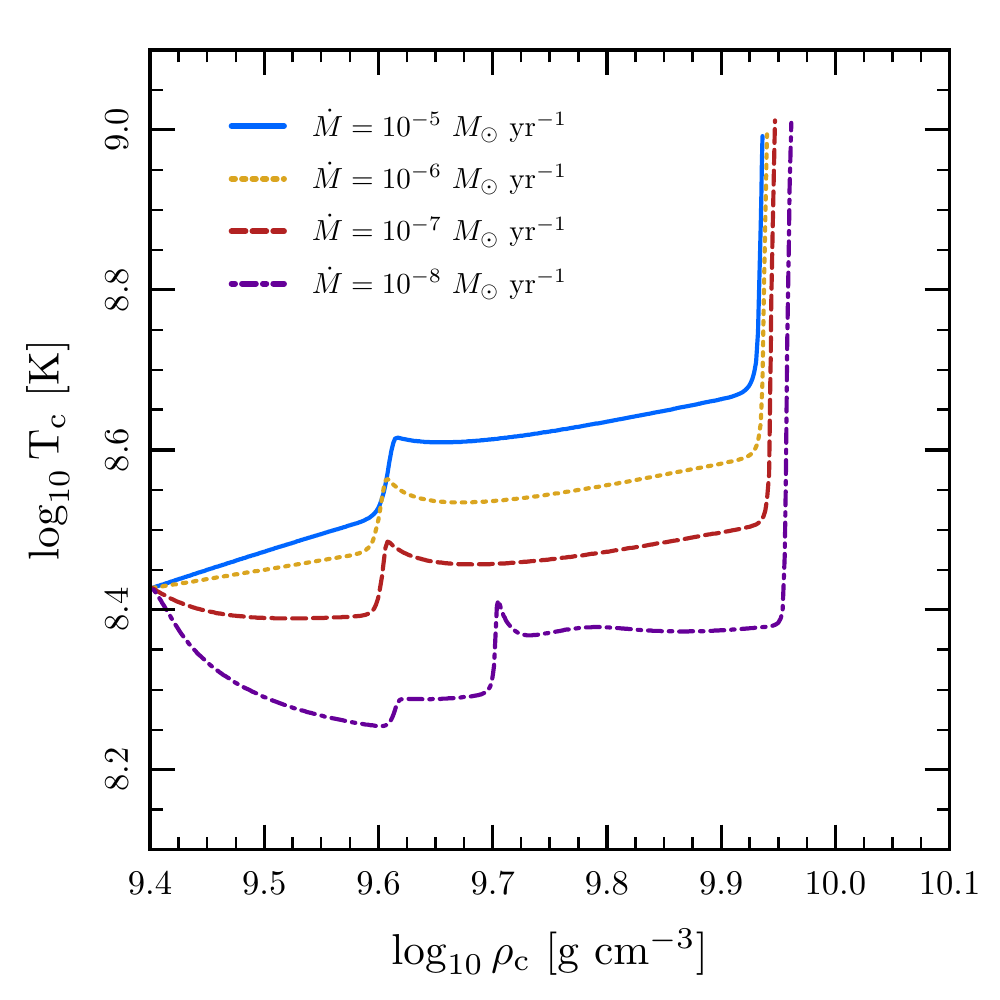}
  \caption{The fiducial model evolved with different values of
    $\dot{M}$.  The central density at which the $A=20$ captures occur
    depends weakly on the accretion rate.  At
    $\dot{M} = 10^{-8} \msunyr$ the central temperature remains low
    enough that the $\magnesium$ and $\sodium[24]$ captures occur at
    two separate critical densities.  The dependence of the oxygen
    ignition density on $\dot{M}$ is weak.}
  \label{fig:mdot}
\end{figure}

In Fig.~\ref{fig:mdot} we show the evolution of our fiducial model
with 4 different accretion rates.  The onset of captures is
less temperature sensitive than one would infer from
Fig.~\ref{fig:critical-capture}.  This is because at a higher
$\dot{M}$, while the quasi-equilibrium core temperature is higher
(increasing the electron capture rates), the compression time is also
shorter, and so the density at which
$t_\mathrm{compress} \approx t_\mathrm{capture}$ ends up having a
weaker dependence on the accretion rate.  At the lowest accretion rate
shown in Fig.~\ref{fig:mdot} ($\dot{M} = 10^{-8} \msunyr$), the
evolution appears qualitatively different.  Looking at
Fig.~\ref{fig:critical-capture}, this is because the central
temperature remains sufficiently low that electron captures on
$\sodium[24]$ do not occur immediately after electron captures on
$\magnesium$, but are delayed until higher densities
($\log_{10} \rho_c \approx 9.7$).

\begin{figure}
  \centering
  \includegraphics[width=\columnwidth]{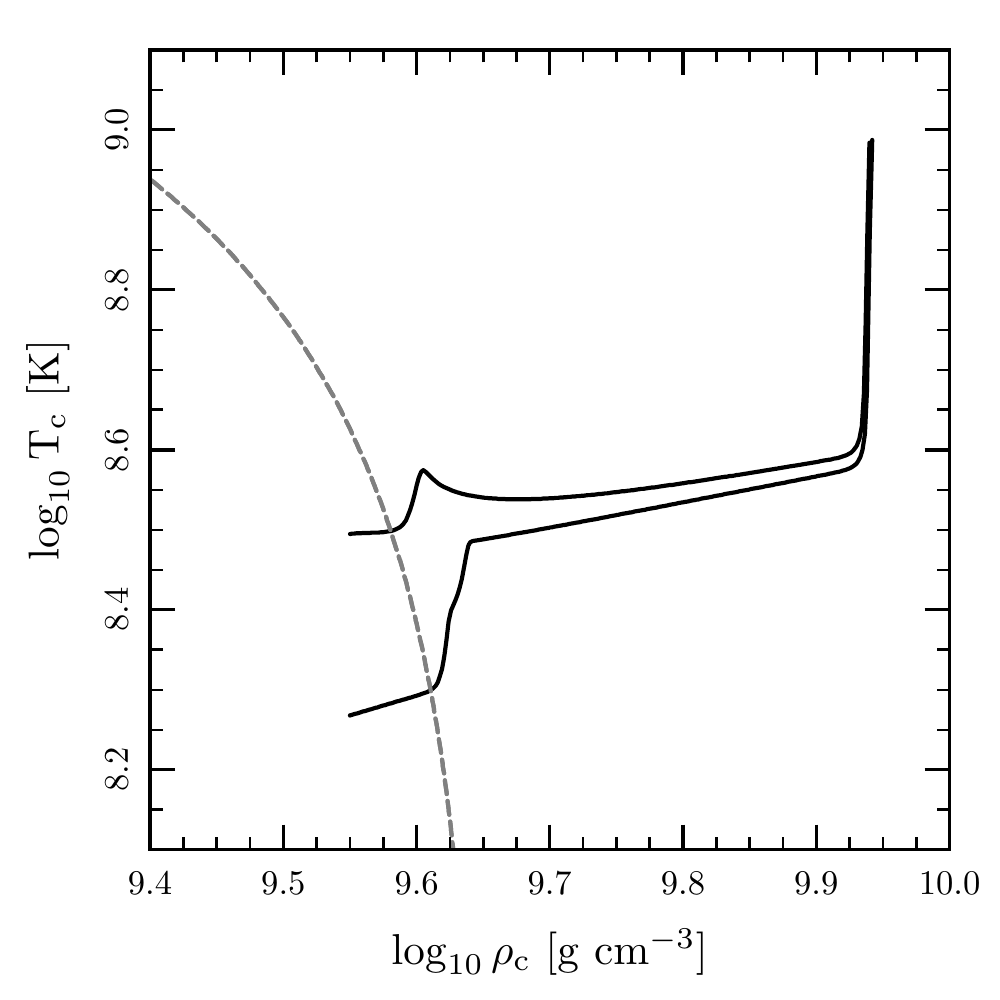}
  \caption{An illustration that the oxygen ignition density is
    independent of the initial central temperature.  The models begin
    with $\log_{10} \rho_c \approx 9.55$, before the $A=24$
    e-captures.  The grey dashed lines show when
    $t_\mathrm{capture} = t_\mathrm{compress}$ for $\magnesium$ and
    the models show the expected temperature dependence for the $A=24$
    captures.  However, by the time the $A=20$ captures occur, the
    temperature differences have been erased by neutrino cooling and
    compressional heating, and thus there is little effect on the
    density at which oxygen ignition occurs.}
  \label{fig:tcore}
\end{figure}

Fig.~\ref{fig:tcore} demonstrates the independence of the oxygen
ignition density on the initial central temperature.  These models
begin right before the $A=24$ captures, at
$\log_{10} \rho_c \approx 9.55$, so that the core temperature does not
change substantially before the onset of the captures.  By the time
the $A=20$ captures occur, the temperature differences have been
erased by neutrino cooling and compressional heating, as the core
evolves towards the $t_\mathrm{compress} = t_\mathrm{cool}$ thermal
state discussed in \S~\ref{sec:analytics}.  As a result, there is
little effect on the density at which oxygen ignition occurs.

\subsection{Effect of a \neon[20] forbidden transition}
\label{sec:parameters-forbidden}

\citet{MartinezPinedo14} discuss the non-unique second forbidden
transition from the $0^+$ ground state of \neon[20] to the $2^+$
ground state of \fluorine[20].  The matrix element for this transition
only has an experimental upper limit.  They show that this transition
can potentially dominate the rate for temperatures less than
$9\times10^8$ K.\footnote{The results of both \citet{MartinezPinedo14}
  and of this work are obtained by treating the phase space factor of
  this second forbidden transition as that of an allowed transition.
  As discussed by \citet{MartinezPinedo14}, the true shape factor
  could contain additional powers of the energy, which would further
  increase the rate, and can potentially offset the possibility that
  the matrix element is below the current experimental upper limit.}

\begin{figure}
  \centering
  \includegraphics[width=\columnwidth]{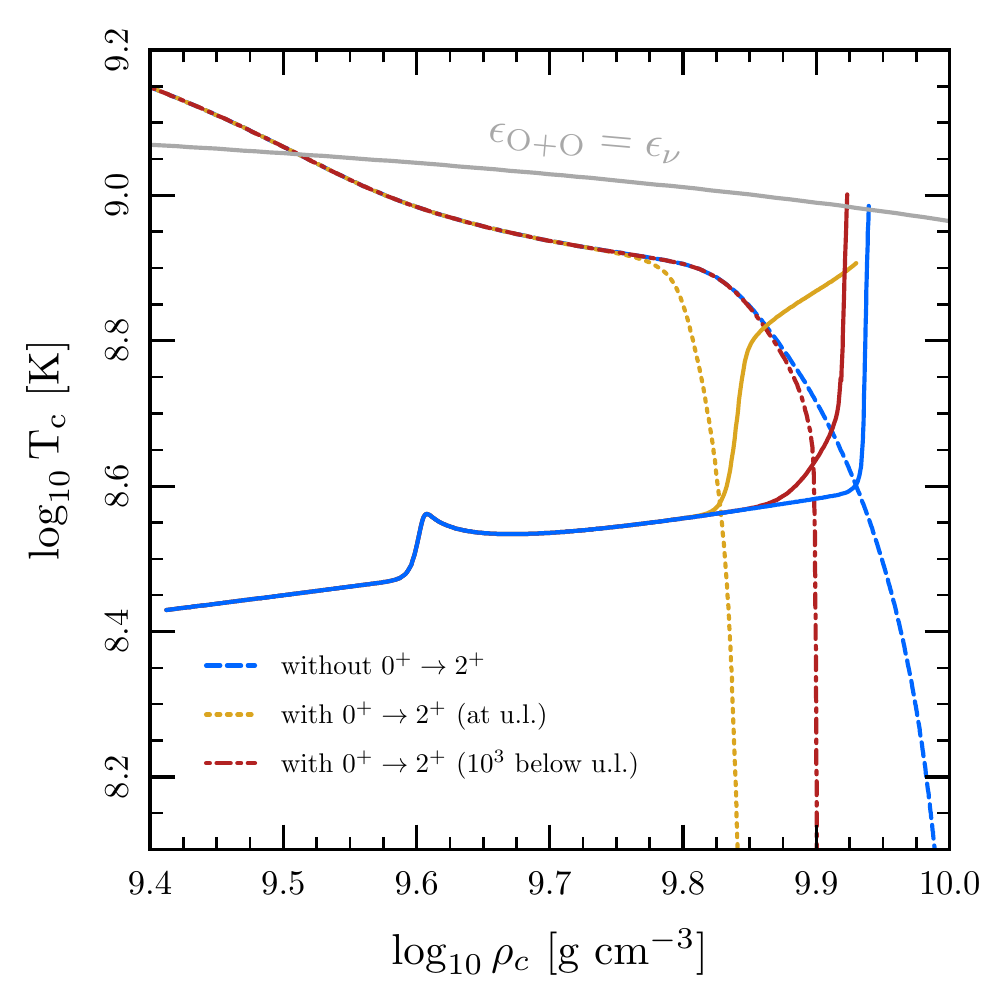}
  \caption{The effect of the second forbidden transition from the
    $0^+$ ground state of \neon[20] to the $2^+$ ground state of
    \fluorine[20]. The dotted, dashed, and dash-dotted lines show
    where the timescale for \neon[20] captures is equal to the
    fiducial compression timescale ($\unit[10^4]{yr}$) for different
    values of the matrix element (see plot legend).  The solid lines
    of matching color show the evolution of the fiducial MESA model
    using these rates.  While the onset of $\neon$ captures shifts
    significantly if the $0^+ \to 2^+$ transition is at the
    experimental upper limit, the shift in the density at which oxygen
    ignition occurs is substantially smaller.}
  \label{fig:ne-capture}
\end{figure}

This transition can affect the critical density at which \neon[20]
captures begin.  The broken lines in Fig.~\ref{fig:ne-capture} shows
the critical curves for \neon[20] capture obtained by setting the
capture rate equal to the fiducial compression rate, corresponding to
$\dot{M} = \unit[10^{-6}]{\msun\,yr^{-1}}$.  With the matrix element at
the current experimental upper limit, the onset of captures is shifted
to lower density (0.15 dex in $\log_{10} \rho$).  At a value a factor
of $10^{3}$ below the upper limit, the shift is very approximately
halved (depending on the temperature).  At a value a factor of
$10^{6}$ below the upper limit, the transition ceases to have a
substantial effect.

The solid lines in Fig.~\ref{fig:ne-capture} show the evolution of our
fiducial model with each of these different choices for the strength
of this transition.  While the choice substantially affects the onset
of $\neon$ captures, it has a less significant effect on the density
for oxygen ignition.  Unlike the other transitions, which reach the
critical capture timescale while they are sub-threshold, this
transition is super-threshold.  Correspondingly, the electron capture
rate is less temperature sensitive.  Its less rapid increase, coupled
with the compression timescale dropping due to the decrease in $Y_e$,
gives time for the core density to increase before the onset of oxygen
ignition.

In the calculation with the transition at the upper limit (solid
yellow line), the central temperature does not reach the oxygen
ignition line.  This is because the thermal structure of the remnant
is such that ignition occurs mildly off-center.  Our future
calculations will determine whether this has any effect on the ensuing
evolution.

\section{Discussion}
\label{sec:discussion}

As described by \citet{Miyaji80}, the final outcome of an ONeMg core
as it approaches the Chandrasekhar mass, either explosion or collapse,
is determined by a competition between the energy release from the
outgoing oxygen deflagration and the energy losses due to electron
captures on the post-deflagration material, which has burned to
nuclear statistical equilibrium (NSE).

As discussed in \S~\ref{sec:analytics}, the small length scale of the
deflagration means that we are unable to follow this phase with the
MESA calculations presented in this paper.  However, in lieu of a full
calculation, we present a few order-of-magnitude estimates relevant to
the outcome.

At the time of collapse, the total energy of our fiducial white dwarf
is $E \approx -\unit[6\times10^{50}]{erg}$.  Oxygen burning to NSE
yields approximately $\unit[1]{MeV}$ per baryon, meaning the energy
release from burning $\unit[0.3]{\msun}$ of material can unbind the
white dwarf.  This energy release is required for the deflagration
wave to significantly change the structure of the star.  Prior to the
deflagration wave burning through $\approx 0.3 \msun$ of material, the
structure of the WD core will remain relatively unchanged unless
electron captures cause collapse.  If the deflagration moves at some
fraction $f$ of the sound speed, the timescale for it to propagate
though the central $0.3 \msun$ is
\begin{equation}
  t_d \approx \int_0^{M_r = 0.3 \msun} \frac{dr}{f c_s} \approx \unit[1]{s}\left(\frac{0.03}{f}\right)~,
  \label{eq:tdef}
\end{equation}
where we have evaluated the integral using the structure of our MESA
model at the end of the calculation.

\citet{Nomoto91} found that the critical deflagration speed that
demarcated the boundary between a model that explodes and a model that
collapses was $f \approx 0.03$.  The work of \citet{Timmes92}, which
simulated conductively-propagating oxygen deflagrations in detail
gives a fitting formula for the laminar deflagration speed of an
oxygen flame of
\begin{equation}
  \label{eq:timmes-speed}
  v_d = \unit[51.8]{km\,s^{-1}}
  \left(\frac{\rho}{\unit[6\times10^{9}]{g\,cm^{-3}}}\right)^{1.06}
  \left(\frac{X_\mathrm{O}}{0.6}\right)^{0.688} ~.
\end{equation}
At $\rho \approx \unit[9\times10^9]{g\,cm^{-3}}$ and
$X_\mathrm{O} = 0.5$, this gives
$v_\mathrm{d} \approx \unit[70]{km\,s^{-1}}$, which corresponds to
$f = v_d/c_s \approx 0.005$. Based on an analysis of the growth of the
Rayleigh-Taylor instability, they conclude that these conductive
flames are likely to remain stable.  Therefore the laminar flame
velocity is representative of the true flame speed in the inner part
of the star.  In particular, see figure 10 in \citet{Timmes92}, noting
that $R(M_r = 0.3 \msun) \approx \unit[300]{km}$.

Based on these flame calculations, as well as several KEPLER
simulations using these speeds, \citet{Timmes92} concluded that above
a core density of $\unit[9\times10^9]{g\,cm^{-3}}$ the white dwarfs
should collapse to a neutron star.  The lowest central density at
which oxygen ignition occurred in our parameter study
(\S~\ref{sec:parameters}) was $\log_{10} \rho_c = 9.93$.  That is
$\rho_c \approx \unit[8.5 \times 10^{9}]{g\,cm^{-3}}$, which is only
marginally below this critical value.

\begin{figure}
  \centering
  \includegraphics[width=\columnwidth]{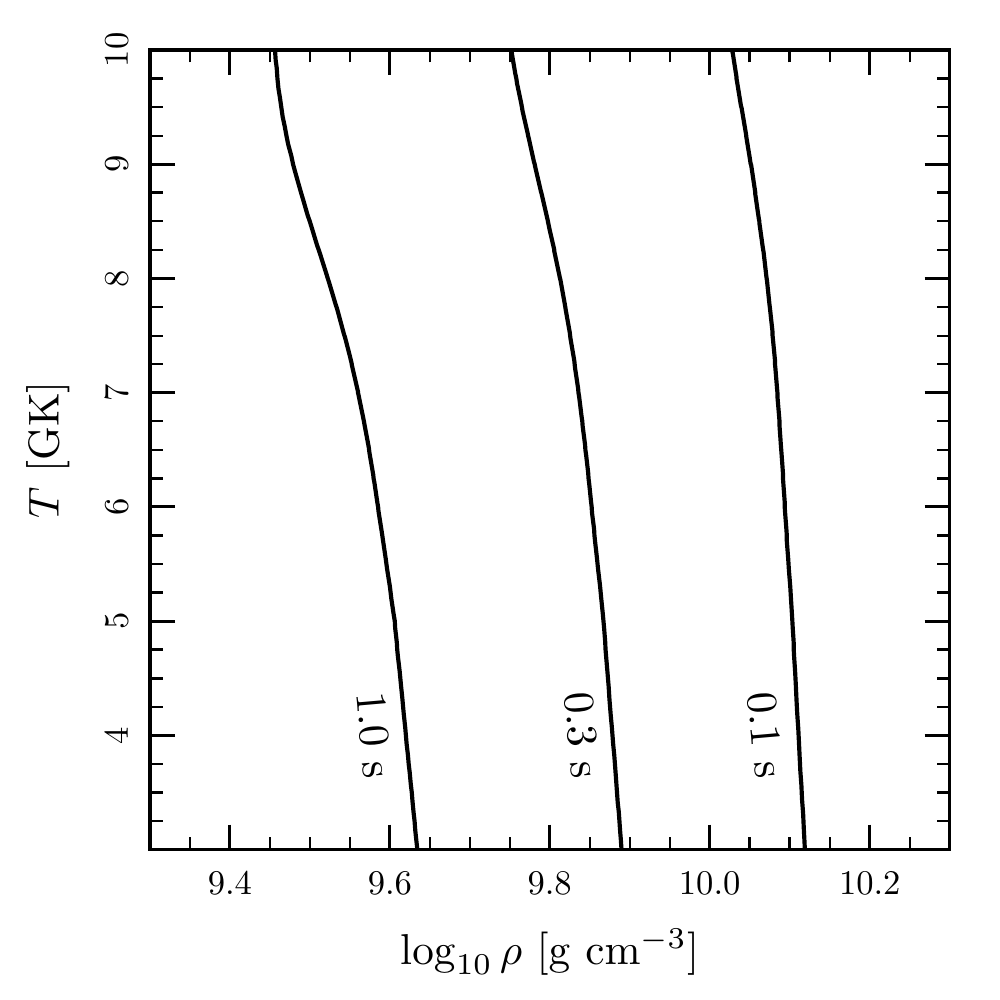}
  \caption{The neutronization timescale for $Y_e = 0.49$, which is
    roughly the central value of $Y_e$ at the end of our fiducial
    calculation.  Contours are labeled by timescale.  This uses the
    NSE electron capture rates from \citet{Seitenzahl09}.  Since our
    compressing ONeMg cores reach oxygen ignition at
    $\log_{10} \rho_c \approx 10$, the relevant neutronization
    timescale is approximately $\unit[0.2]{s}$.  This is less than the
    timescale for the O deflagration wave to release enough energy to
    unbind the star (equation~\ref{eq:tdef}), suggesting that
    collapse to a NS will ensue.}
  \label{fig:tneut}
\end{figure}

The timescale on which the core is neutronizing due to electron
captures on the NSE-composition material can be written as
$t_n = (d \ln Y_e/dt)^{-1}$. The methods presented in this paper are
not appropriate for calculating weak rates in NSE material.  Instead,
we take $\dot{Y_e}$ from tables generated by \citet{Seitenzahl09}.
Fig.~\ref{fig:tneut} shows the neutronization timescale as a function
of density and temperature for $Y_e = 0.49$, the approximate central
value in our fiducial model at oxygen ignition.
Once the deflagration forms, the density of the post-deflagration
material is less than the cold, upstream material.  The MESA models
reach oxygen ignition at $\log_{10} \rho_c \approx 10$, where this
density change is small, $\Delta \rho / \rho \approx 0.1$
\citep{Timmes92}.  Therefore, the density of the post-deflagration ash
will be approximately the same as the density at which oxygen ignites,
so Fig.~\ref{fig:tneut} indicates that the relevant neutronization
timescale is approximately $\unit[0.2]{s}$.

This estimate of the neutronization timescale is sufficiently shorter
than the timescale on which the deflagration wave unbinds the star
(equation~\ref{eq:tdef}) that it suggests that the end
result of oxygen ignition following e-captures on $\neon$ will be
collapse to a NS rather than a thermonuclear explosion.  Future work
will clarify this in the context of full-star simulations.

\section{Conclusions}
\label{sec:conclusions}

We have provided an updated analytic and numerical understanding of
the evolution of accreting and compressing ONeMg cores up to the
initiation of oxygen burning in the core. This study was enabled by
new capabilities of the MESA \citep{Paxton11,Paxton13,Paxton15} stellar
evolution code.  In particular, we have implemented a highly accurate
treatment of the key electron capture rates (on $A = 20$ and $24$
nuclei) using modern microphysics from \citet{MartinezPinedo14}, which
is summarized in Appendix~\ref{sec:ecapture}.

We have demonstrated analytically and numerically that neither
$^{24}$Mg or $^{20}$Ne captures release sufficient heat to generate
convection in the core.  Instead, the core undergoes a thermal runaway
triggered by the energy released by $^{20}$Ne captures. This centrally
concentrated runaway initiates oxygen burning and launches an outgoing
oxygen deflagration wave at a time when the central density is at
least $\unit[8.5 \times 10^{9}]{g\,cm^{-3}}$.  Based on order of
magnitude estimates and previous work of \citet{Timmes92}, we expect
objects which ignite oxygen at such high densities will collapse and
form a neutron star due to continued electron captures on the NSE
ashes produced by oxygen burning.

Given the sensitivity of the final outcome of compressing ONeMg cores
to the central density at the time oxygen burning begins (see \S
\ref{sec:discussion}), we also performed a parameter study which
demonstrated the influence of a number of factors on this density.  We
investigated the effects of varying the initial $\magnesium$ fraction
(\S~\ref{sec:parameters-xmg}), the initial central temperature and
accretion rate (\S~\ref{sec:parameters-tc-mdot}), as well as the
potential inclusion of a particular forbidden transition
(\S~\ref{sec:parameters-forbidden}).
Figures~\ref{fig:compression}~and~\ref{fig:critical_xmg} demonstrate
that values of $X_\mathrm{Mg} \ga 0.07$ cause the core to contract
more rapidly after $A=24$ the captures, which leads to oxygen ignition
at higher densities (thus further favoring collapse to form a neutron
star).  We also demonstrated the importance of the balance between
neutrino cooling and compressional heating in setting the central
temperature of ONeMg cores during much of their evolution
\citep{Paczynski71}.  This implies that the core typically loses
memory of its initial central temperature.

The approach of an ONeMg core to the Chandrasekhar mass is relevant to
the late stages of evolution for super asymptotic giant branch stars,
binary systems with an accreting ONeMg WD, and the remnants of WD-WD
mergers. Our calculations here are an important step in producing more
realistic progenitor models for these studies.

\section*{Acknowledgments}
We thank Gabriel Mart\'{i}nez-Pinedo and Christopher Sullivan for
discussing their work in \citet{MartinezPinedo14} with us in advance
of publication.  We thank Jared Brooks, Dan Kasen, Christian Ott,
Sterl Phinney and Ken Shen for useful discussions.  We thank Frank
Timmes for his comments on the manuscript.  We acknowledge stimulating
workshops at Sky House where these ideas germinated. JS is supported
by the National Science Foundation Graduate Research Fellowship
Program under Grant No.~DGE 1106400 and by NSF grant AST-1205732.  EQ
is supported in part by a Simons Investigator award from the Simons
Foundation and the David and Lucile Packard Foundation.  LB is
supported by the National Science Foundation under grants PHY
11-25915, AST 11-09174, and AST 12-05574.  This research used the
SAVIO computational cluster resource provided by the Berkeley Research
Computing program at the University of California Berkeley (Supported
by UC Chancellor, UC Berkeley Vice Chancellor of Research and Office
of the CIO).  This research has made use of NASA's Astrophysics Data
System.

\bibliography{ONeMgCores}

\begin{thebibliography}{}

\bibitem[\protect\citeauthoryear{{Aparicio}}{{Aparicio}}{1998}]{Aparicio98}
{Aparicio} J.~M.,  1998, \apjs, 117, 627

\bibitem[\protect\citeauthoryear{{Canal}, {Isern} \& {Labay}}{{Canal}
  et~al.}{1992}]{Canal92}
{Canal} R.,  {Isern} J.,    {Labay} J.,  1992, \apjl, 398, L49

\bibitem[\protect\citeauthoryear{{Chabrier} \& {Potekhin}}{{Chabrier} \&
  {Potekhin}}{1998}]{Chabrier98}
{Chabrier} G.,  {Potekhin} A.~Y.,  1998, \pre, 58, 4941

\bibitem[\protect\citeauthoryear{{Commins}}{{Commins}}{1973}]{Commins73}
{Commins} E.,  1973, {Weak Interactions}.
McGraw-Hill

\bibitem[\protect\citeauthoryear{{Couch} \& {Loumos}}{{Couch} \&
  {Loumos}}{1974}]{Couch74}
{Couch} R.~G.,  {Loumos} G.~L.,  1974, \apj, 194, 385

\bibitem[\protect\citeauthoryear{{Cox}}{{Cox}}{1968}]{Cox68}
{Cox} J.~P.,  1968, {Principles of stellar structure - Vol.1: Physical
  principles; Vol.2: Applications to stars}

\bibitem[\protect\citeauthoryear{{Dewitt}, {Graboske} \& {Cooper}}{{Dewitt}
  et~al.}{1973}]{Dewitt73}
{Dewitt} H.~E.,  {Graboske} H.~C.,    {Cooper} M.~S.,  1973, \apj, 181, 439

\bibitem[\protect\citeauthoryear{{Farmer}, {Fields} \& {Timmes}}{{Farmer}
  et~al.}{2015}]{Farmer15}
{Farmer} R.,  {Fields} C.~E.,    {Timmes} F.~X.,  2015, \apj, 807, 184

\bibitem[\protect\citeauthoryear{{Firestone}}{{Firestone}}{2007}]{Firestone07b}
{Firestone} R.~B.,  2007, Nuclear Data Sheets, 108, 2319

\bibitem[\protect\citeauthoryear{{Fuller}, {Fowler} \& {Newman}}{{Fuller}
  et~al.}{1980}]{Fuller80}
{Fuller} G.~M.,  {Fowler} W.~A.,    {Newman} M.~J.,  1980, \apjs, 42, 447

\bibitem[\protect\citeauthoryear{{Fuller}, {Fowler} \& {Newman}}{{Fuller}
  et~al.}{1985}]{Fuller85}
{Fuller} G.~M.,  {Fowler} W.~A.,    {Newman} M.~J.,  1985, \apj, 293, 1

\bibitem[\protect\citeauthoryear{{Gove} \& {Martin}}{{Gove} \&
  {Martin}}{1971}]{Gove71}
{Gove} N.~B.,  {Martin} M.~J.,  1971, Atomic Data and Nuclear Data Tables, 10,
  205

\bibitem[\protect\citeauthoryear{{Guti{\'e}rrez}, {Canal} \&
  {Garc{\'{\i}}a-Berro}}{{Guti{\'e}rrez} et~al.}{2005}]{Gutierrez05}
{Guti{\'e}rrez} J.,  {Canal} R.,    {Garc{\'{\i}}a-Berro} E.,  2005, \aap, 435,
  231

\bibitem[\protect\citeauthoryear{{Gutierrez}, {Garcia-Berro}, {Iben} Jr.,
  {Isern}, {Labay} \& {Canal}}{{Gutierrez} et~al.}{1996}]{Gutierrez96}
{Gutierrez} J.,  {Garcia-Berro} E.,  {Iben} Jr. I.,  {Isern} J.,  {Labay} J.,
   {Canal} R.,  1996, \apj, 459, 701

\bibitem[\protect\citeauthoryear{{Ichimaru}}{{Ichimaru}}{1993}]{Ichimaru93}
{Ichimaru} S.,  1993, Reviews of Modern Physics, 65, 255

\bibitem[\protect\citeauthoryear{{Isern}, {Canal} \& {Labay}}{{Isern}
  et~al.}{1991}]{Isern91}
{Isern} J.,  {Canal} R.,    {Labay} J.,  1991, \apjl, 372, L83

\bibitem[\protect\citeauthoryear{{Itoh}, {Tomizawa}, {Tamamura}, {Wanajo} \&
  {Nozawa}}{{Itoh} et~al.}{2002}]{Itoh02}
{Itoh} N.,  {Tomizawa} N.,  {Tamamura} M.,  {Wanajo} S.,    {Nozawa} S.,  2002,
  \apj, 579, 380

\bibitem[\protect\citeauthoryear{{Jones}, {Hirschi} \& {Nomoto}}{{Jones}
  et~al.}{2014}]{Jones14}
{Jones} S.,  {Hirschi} R.,    {Nomoto} K.,  2014, \apj, 797, 83

\bibitem[\protect\citeauthoryear{{Jones}, {Hirschi}, {Nomoto}, {Fischer},
  {Timmes}, {Herwig}, {Paxton}, {Toki}, {Suzuki}, {Mart{\'{\i}}nez-Pinedo},
  {Lam} \& {Bertolli}}{{Jones} et~al.}{2013}]{Jones13}
{Jones} S.,  {Hirschi} R.,  {Nomoto} K.,  {Fischer} T.,  {Timmes} F.~X.,
  {Herwig} F.,  {Paxton} B.,  {Toki} H.,  {Suzuki} T.,
  {Mart{\'{\i}}nez-Pinedo} G.,  {Lam} Y.~H.,    {Bertolli} M.~G.,  2013, \apj,
  772, 150

\bibitem[\protect\citeauthoryear{{Juodagalvis}, {Langanke}, {Hix},
  {Mart{\'{\i}}nez-Pinedo} \& {Sampaio}}{{Juodagalvis}
  et~al.}{2010}]{Juodagalvis10}
{Juodagalvis} A.,  {Langanke} K.,  {Hix} W.~R.,  {Mart{\'{\i}}nez-Pinedo} G.,
   {Sampaio} J.~M.,  2010, Nuclear Physics A, 848, 454

\bibitem[\protect\citeauthoryear{{Langer}}{{Langer}}{1991}]{Langer91}
{Langer} N.,  1991, \aap, 252, 669

\bibitem[\protect\citeauthoryear{{Langer}, {Fricke} \& {Sugimoto}}{{Langer}
  et~al.}{1983}]{Langer83}
{Langer} N.,  {Fricke} K.~J.,    {Sugimoto} D.,  1983, \aap, 126, 207

\bibitem[\protect\citeauthoryear{{Mart{\'{\i}}nez-Pinedo}, {Lam}, {Langanke},
  {Zegers} \& {Sullivan}}{{Mart{\'{\i}}nez-Pinedo}
  et~al.}{2014}]{MartinezPinedo14}
{Mart{\'{\i}}nez-Pinedo} G.,  {Lam} Y.~H.,  {Langanke} K.,  {Zegers} R.~G.~T.,
    {Sullivan} C.,  2014, \prc, 89, 045806

\bibitem[\protect\citeauthoryear{{Miyaji} \& {Nomoto}}{{Miyaji} \&
  {Nomoto}}{1987}]{Miyaji87}
{Miyaji} S.,  {Nomoto} K.,  1987, \apj, 318, 307

\bibitem[\protect\citeauthoryear{{Miyaji}, {Nomoto}, {Yokoi} \&
  {Sugimoto}}{{Miyaji} et~al.}{1980}]{Miyaji80}
{Miyaji} S.,  {Nomoto} K.,  {Yokoi} K.,    {Sugimoto} D.,  1980, \pasj, 32, 303

\bibitem[\protect\citeauthoryear{{Nomoto}}{{Nomoto}}{1984}]{Nomoto84a}
{Nomoto} K.,  1984, \apj, 277, 791

\bibitem[\protect\citeauthoryear{{Nomoto} \& {Kondo}}{{Nomoto} \&
  {Kondo}}{1991}]{Nomoto91}
{Nomoto} K.,  {Kondo} Y.,  1991, \apjl, 367, L19

\bibitem[\protect\citeauthoryear{{Oda}, {Hino}, {Muto}, {Takahara} \&
  {Sato}}{{Oda} et~al.}{1994}]{Oda94}
{Oda} T.,  {Hino} M.,  {Muto} K.,  {Takahara} M.,    {Sato} K.,  1994, Atomic
  Data and Nuclear Data Tables, 56, 231

\bibitem[\protect\citeauthoryear{{Paczy{\'n}ski}}{{Paczy{\'n}ski}}{1971}]{Paczynski71}
{Paczy{\'n}ski} B.,  1971, \actaa, 21, 271

\bibitem[\protect\citeauthoryear{{Paxton}, {Bildsten}, {Dotter}, {Herwig},
  {Lesaffre} \& {Timmes}}{{Paxton} et~al.}{2011}]{Paxton11}
{Paxton} B.,  {Bildsten} L.,  {Dotter} A.,  {Herwig} F.,  {Lesaffre} P.,
  {Timmes} F.,  2011, \apjs, 192, 3

\bibitem[\protect\citeauthoryear{{Paxton}, {Cantiello}, {Arras}, {Bildsten},
  {Brown}, {Dotter}, {Mankovich}, {Montgomery}, {Stello}, {Timmes} \&
  {Townsend}}{{Paxton} et~al.}{2013}]{Paxton13}
{Paxton} B.,  {Cantiello} M.,  {Arras} P.,  {Bildsten} L.,  {Brown} E.~F.,
  {Dotter} A.,  {Mankovich} C.,  {Montgomery} M.~H.,  {Stello} D.,  {Timmes}
  F.~X.,    {Townsend} R.,  2013, \apjs, 208, 4

\bibitem[\protect\citeauthoryear{{Paxton}, {Marchant}, {Schwab}, {Bauer},
  {Bildsten}, {Cantiello}, {Dessart}, {Farmer}, {Hu}, {Langer}, {Townsend},
  {Townsley} \& {Timmes}}{{Paxton} et~al.}{2015}]{Paxton15}
{Paxton} B.,  {Marchant} P.,  {Schwab} J.,  {Bauer} E.~B.,  {Bildsten} L.,
  {Cantiello} M.,  {Dessart} L.,  {Farmer} R.,  {Hu} H.,  {Langer} N.,
  {Townsend} R.~H.~D.,  {Townsley} D.~M.,    {Timmes} F.~X.,  2015, ArXiv
  e-prints

\bibitem[\protect\citeauthoryear{{Potekhin} \& {Chabrier}}{{Potekhin} \&
  {Chabrier}}{2000}]{Potekhin00}
{Potekhin} A.~Y.,  {Chabrier} G.,  2000, \pre, 62, 8554

\bibitem[\protect\citeauthoryear{{Potekhin}, {Chabrier} \& {Rogers}}{{Potekhin}
  et~al.}{2009}]{Potekhin09a}
{Potekhin} A.~Y.,  {Chabrier} G.,    {Rogers} F.~J.,  2009, \pre, 79, 016411

\bibitem[\protect\citeauthoryear{{Ritossa}, {Garc{\'{\i}}a-Berro} \& {Iben}
  Jr.}{{Ritossa} et~al.}{1999}]{Ritossa99}
{Ritossa} C.,  {Garc{\'{\i}}a-Berro} E.,    {Iben} Jr. I.,  1999, \apj, 515,
  381

\bibitem[\protect\citeauthoryear{{Saio} \& {Nomoto}}{{Saio} \&
  {Nomoto}}{1985}]{Saio85}
{Saio} H.,  {Nomoto} K.,  1985, \aap, 150, L21

\bibitem[\protect\citeauthoryear{{Schwab}, {Shen}, {Quataert}, {Dan} \&
  {Rosswog}}{{Schwab} et~al.}{2012}]{Schwab12}
{Schwab} J.,  {Shen} K.~J.,  {Quataert} E.,  {Dan} M.,    {Rosswog} S.,  2012,
  \mnras, 427, 190

\bibitem[\protect\citeauthoryear{{Seitenzahl}, {Townsley}, {Peng} \&
  {Truran}}{{Seitenzahl} et~al.}{2009}]{Seitenzahl09}
{Seitenzahl} I.~R.,  {Townsley} D.~M.,  {Peng} F.,    {Truran} J.~W.,  2009,
  Atomic Data and Nuclear Data Tables, 95, 96

\bibitem[\protect\citeauthoryear{{Shapiro} \& {Teukolsky}}{{Shapiro} \&
  {Teukolsky}}{1983}]{Shapiro83}
{Shapiro} S.~L.,  {Teukolsky} S.~A.,  1983, {Black holes, white dwarfs, and
  neutron stars: The physics of compact objects}

\bibitem[\protect\citeauthoryear{{Shen}, {Bildsten}, {Kasen} \&
  {Quataert}}{{Shen} et~al.}{2012}]{Shen12}
{Shen} K.~J.,  {Bildsten} L.,  {Kasen} D.,    {Quataert} E.,  2012, \apj, 748,
  35

\bibitem[\protect\citeauthoryear{{Takahara}, {Hino}, {Oda}, {Muto}, {Wolters},
  {Glaudemans} \& {Sato}}{{Takahara} et~al.}{1989}]{Takahara89}
{Takahara} M.,  {Hino} M.,  {Oda} T.,  {Muto} K.,  {Wolters} A.~A.,
  {Glaudemans} P.~W.~M.,    {Sato} K.,  1989, Nuclear Physics A, 504, 167

\bibitem[\protect\citeauthoryear{{Takahashi}, {Yoshida} \& {Umeda}}{{Takahashi}
  et~al.}{2013}]{Takahashi13}
{Takahashi} K.,  {Yoshida} T.,    {Umeda} H.,  2013, \apj, 771, 28

\bibitem[\protect\citeauthoryear{{Tilley}, {Cheves}, {Kelley}, {Raman} \&
  {Weller}}{{Tilley} et~al.}{1998}]{Tilley98}
{Tilley} D.~R.,  {Cheves} C.~M.,  {Kelley} J.~H.,  {Raman} S.,    {Weller}
  H.~R.,  1998, Nuclear Physics A, 636, 249

\bibitem[\protect\citeauthoryear{{Timmes} \& {Swesty}}{{Timmes} \&
  {Swesty}}{2000}]{Timmes00b}
{Timmes} F.~X.,  {Swesty} F.~D.,  2000, \apjs, 126, 501

\bibitem[\protect\citeauthoryear{{Timmes} \& {Woosley}}{{Timmes} \&
  {Woosley}}{1992}]{Timmes92}
{Timmes} F.~X.,  {Woosley} S.~E.,  1992, \apj, 396, 649

\bibitem[\protect\citeauthoryear{{Yakovlev} \& {Shalybkov}}{{Yakovlev} \&
  {Shalybkov}}{1989}]{Yakovlev89}
{Yakovlev} D.~G.,  {Shalybkov} D.~A.,  1989, Astrophysics and Space Physics
  Reviews, 7, 311

\bibitem[\protect\citeauthoryear{{Yoon}, {Langer} \& {Norman}}{{Yoon}
  et~al.}{2006}]{Yoon06}
{Yoon} S.-C.,  {Langer} N.,    {Norman} C.,  2006, \aap, 460, 199

\end{thebibliography}

\appendix

\section{Physics of electron-capture and beta-decay}
\label{sec:ecapture}

We are interested in the electron-capture reaction
\begin{equation}
  \label{eq:electron-capture}
  (Z, N) + e^- \to (Z-1,N+1)+ \nu_e
\end{equation}
and its reverse reaction, $\beta$-decay
\begin{equation}
  \label{eq:beta-decat}
  (Z,N) \to (Z+1, N-1) + e^- + \bar{\nu}_e
\end{equation}
where $Z$ and $N$ are respectively the proton and neutron number of
the nucleus.  For nuclei in a dense plasma where the electrons are
degenerate, the rates of these processes can depend sensitively on the
density (though the electron distribution function) and temperature
(though the occupation of nuclear energy levels and the electron
distribution function).  The neutrinos are able to free stream out of
the star, and therefore neutrino phase-space is effectively unfilled.
For a more thorough discussion of the physics of weak reactions in
stellar environments, see e.g. \citet{Fuller80,Fuller85}.

This section summarizes a simple framework for the rates of these weak
processes.  More detailed microscopic calculations exist in the
literature such as those presented in \citet{Oda94}.  However, those
particular tables are sufficiently sparse that numerical
considerations related to interpolation cause us to elect to use rates
calculated in the manner described here, rather than
interpolate in tables from more detailed calculations.

The rate of the electron capture or $\beta$-decay transition from the
$i$-th state of the parent nucleus to the $j$-th state of the daughter
nucleus can be written as \citep[e.g.][]{Fuller80}
\begin{equation}
  \label{eq:rate}
  \lambda_{ij} = \frac{\ln 2}{(ft)_{ij}} I(\mu, T, Q_{ij}),
\end{equation}
where $(ft)$ is the comparative half-life and can be either measured
experimentally or theoretically calculated from the weak-interaction
nuclear matrix elements.  I is a phase space factor which depends on
the temperature $T$, electron chemical potential $\mu$, and the energy
difference $Q_{ij}$ between the (i-th) parent and (j-th) daughter
nuclear states.
\begin{equation}
  \label{eq:q-mu}
  Q_{ij} = \left(\mu_p - \mu_d\right) + E_i - E_j~,
\end{equation}
where $\mu_p$ and $\mu_d$ are the chemical potentials of the nuclei.
For a classical ideal gas, the chemical potential is
\begin{equation}
  \label{eq:ideal-gas-mu}
  \mu_I = m_Ic^2 + kT \ln \left(\frac{n_I}{n_q}\right)~,
\end{equation}
where $m_I$ is the rest mass, $n_I$ is the number density, and $n_q =
(2 \pi m_I k T / h^2)^{3/2}$.  Therefore,
\begin{equation}
  \label{eq:q-expanded}
  Q_{ij} = \left(M_p - M_d\right) c^2 + kT \ln \left(\frac{n_p}{n_d} \right) + E_i - E_j
\end{equation}
where $M_p$ and $M_d$ are the nuclear rest masses of the parent and
daughter nuclei, respectively. Since
$\left|M_p - M_d\right| c^2 \approx \unit[5]{MeV}$ for the isotopes we
consider and we restrict ourselves to temperatures
$T < \unit[10^9]{K}$ (so $kT < \unit[100]{keV}$), the term
$kT \ln \left(\frac{n_p}{n_d} \right)$ is negligible in comparison and
we discard it, leaving
\begin{equation}
  \label{eq:q}
  Q_{ij} = \left(M_p - M_d\right) c^2 + E_i - E_j ~.
\end{equation}
Though not generally true, for the set of transitions that we
consider, this definition means that $Q_{ij}<0$ for e-capture and
$Q_{ij}>0$ for $\beta$-decay.

We work in the allowed approximation, which neglects all total lepton
angular momentum ($L = 0$).  This restricts us to the following
transitions and corresponding selection rules
\citep[e.g.][]{Commins73}: Fermi transitions, where the total lepton
spin is $S = 0$, and therefore the initial and final nuclear spins are
equal ($J_i = J_j$), and Gamow-Teller transitions, where $S = 1$, and
therefore $J_i = J_j, J_j \pm 1$ (excluding $J_i = J_f = 0$). In both
cases, these is no parity change: $\pi_i \pi_f = +1$.

At low temperature, the electron chemical potential is approximately the Fermi
energy $E_F$ (the first correction enters at order $(kT/E_F)^2$), and
so we use the terms Fermi energy and chemical potential
interchangeably.  In the relativistic limit, the chemical potential
can be approximated as
\begin{equation}
  \label{eq:chemical-potential}
  \mu \approx E_F = 5.16 \left( \frac{\rho Y_e}{10^9 \mathrm{g\, cm^{-3}}}\right)^{1/3} \text{MeV},
\end{equation}
where $Y_e = \sum_i Z_i X_i / A_i$ is the electron fraction. $Z_i$,
$X_i$, and $A_i$ are respectively the charge, mass fraction, and
atomic mass of the i-th species.

The total rate of the process is the sum of the individual transition
rates from the $i$-th parent state to the $j$-th daughter state,
$\lambda_{ij}$, weighted by the occupation probability of the $i$-th
parent state, $p_i$.
\begin{equation}
  \label{eq:15}
  \lambda_{\mathrm{total}} = \sum_{i} p_i \sum_{j} \lambda_{ij},
\end{equation}
The $i$-sum is over all parent states and the $j$-sum is over all
daughter states.  We will always assume that the nuclear states are
populated with a thermal (Boltzmann) distribution.  Some parent nuclei
preferentially capture into excited daughter states, but these excited
states decay via $\gamma$-ray emission with a typical timescale
$\sim 10^{-12}$ s.  Therefore the level population returns to a
thermal distribution on a timescale much shorter than the evolutionary
timescales of interest.\footnote{There is one case in which this
  hierarchy of timescales is not so obvious.  The first excited state
  of $\sodium[24]$ is metastable with a half-life of
  $2 \times 10^{-2}$ s \citep{Firestone07b} and ground state of
  $\magnesium[24]$ preferentially captures into this excited state.
  If the capture rate from this excited state to $\neon[24]$ were
  approximately equal or greater than the rate of decay via
  $\gamma$-ray emission, the relative state populations would be
  effectively non-thermal.  Using the parameters of this transition as
  listed in Table \ref{tab:transitions}, the capture timescale is
  approximately equal to the half-life at a critical density of
  $\log_{10} \rho \approx 10.5$ (for $Y_e \approx 0.5$), which is
  safely outside of the density range that we consider in this work.}
The occupation probability is
\begin{equation}
  \label{eq:occupation-probability}
  p_i = \frac{2 J_i + 1}{P(T)} \exp\left(-\beta E_i\right)~,
\end{equation}
where $P(T)$, the nuclear partition function is
$P(T) = \sum_i (2 J_i + 1) e^{-\beta E_i}$ and we define
$\beta = (kT)^{-1}$.

The remainder of this section considers the rate of a single allowed
($L=0$) transition in detail, and so for convenience we drop the
$i$,$j$ subscripts.  In the case of electron capture, the phase space
factor is \citep[e.g.,][]{Fuller80}
\begin{equation}
  \label{eq:Iec}
  I_{\mathrm{ec}} = \frac{1}{(m_ec^2)^5}\int_{-Q}^{\infty} \frac{E_e^2 E_\nu^2}{1 + \exp[\beta (E_e - \mu)]} G(Z,E_e) dE_e~,
\end{equation}
along with the energy conservation relationship $E_e + Q = E_\nu$. The
quantity $G$ is defined as
\begin{equation}
  G(Z,E_e) = \frac{\sqrt{E_e^2 - (m_e c^2)^2}}{E_e} F(Z,E_e)
\end{equation}
where $F(Z,E_e)$ is the relativistic Coulomb barrier factor
\citep{Gove71}.  We make the approximation that the electrons are
relativistic.  In this limit,
\begin{equation}
  G(Z,E_e) \approx \left(\frac{4 \pi E_e R}{hc}\right)^{-\alpha^2 Z^2} \exp\left(\pi\alpha Z\right)~,
\end{equation}
where $\alpha$ is the fine structure constant and $R$ is the size of
the nucleus \citep{Fuller80}.  We are considering nuclei with $Z
\approx 10$, $A \approx 20$ (and so $R \approx \unit[3]{fm}$), at
densities such that $E_e \approx \unit[5]{MeV}$.  Therefore, the value
of the first term is $\left(\frac{4 \pi E_e R}{\hbar
    c}\right)^{-\alpha^2 Z^2} \approx 0.999$, with an extremely weak
$E_e$ and $R$ dependence (since $\alpha^2Z^2 \approx 0.005$).
Therefore we treat $G(Z,E_e)$ as a constant with a value of $\exp(\pi
\alpha Z)$.

Changing to dimensionless variables $\epsilon \equiv
\frac{E}{m_ec^2} $ and $q \equiv \frac{Q}{m_ec^2}$, the phase space
integral becomes
\begin{equation}
  I_{\mathrm{ec}} = e^{\pi\alpha Z} \int_{-q}^{\infty}
  \frac{\epsilon^2 (\epsilon + q)^2}
  {1 + \exp[\beta m_e c^2(\epsilon - \mu)]} d\epsilon ~~~.
\end{equation}
We rewrite this integral in
simpler form as
\begin{equation}
  \label{eq:Iec-fd}
  I_{\mathrm{ec}} = \frac{e^{\pi\alpha Z}}{(\beta m_e c^2)^5}\left[ F_4(\eta+\zeta) - 2 \zeta F_3(\eta+\zeta) + \zeta^2 F_2(\eta+\zeta) \right]~,
\end{equation}
where we have defined the quantities $\eta = \beta \mu$ and $\zeta = \beta q$ and
\begin{equation}
  \label{eq:fdintegral}
  F_k(y) = \int_0^{\infty} \frac{x^k}{1 + \exp(x-y)} dx~~~,
\end{equation}
is the complete Fermi integral.  Evaluating the rate requires
evaluating three complete Fermi integrals, for which efficient
numerical routines exist \citep[e.g.][]{Aparicio98}.

In addition to the e-capture rate, we need the rate of energy loss via neutrinos, which for a single transition can be
written as
\begin{equation}
  \varepsilon_{\nu, ij} = \frac{ m_e c^2 \ln 2}{(ft)_{ij}} J(\mu, T, Q_{ij})~,
\end{equation}
where $J$ is phase space factor, defined by an integral similar to
equation~\eqref{eq:Iec}, except with an additional power of the neutrino
energy:
\begin{equation}
  \label{eq:Jec}
  J_{\mathrm{ec}} = \frac{1}{(m_ec^2)^6}\int_{-Q}^{\infty} \frac{E_e^2 E_\nu^3}{1 + \exp[\beta (E_e - \mu)]} G(Z,E_e) dE_e ~.
\end{equation}
In terms of complete Fermi integrals, this is
\begin{equation}
  \label{eq:Jec-fd}
  J_{\mathrm{ec}} = \frac{e^{\pi\alpha Z}}{(\beta m_e c^2)^{6}} \left[ F_5(\eta + \zeta) - 2 \zeta F_4(\eta+\zeta) + \zeta^2 F_3(\eta+\zeta) \right]~.
\end{equation}
The total neutrino loss rate can calculated via an occupation-weighted
average, analogous to that used to calculated the total rate in
equation \eqref{eq:15}.

In the case of $\beta$-decay, the phase space factor is
\begin{equation}
  \label{eq:Ibeta}
  I_{\beta} = \int_{m_ec^2}^{Q} \frac{E_e^2 E_\nu^2}{1 + \exp[-\beta (E - \mu)]} G(Z,E) dE_e~,
\end{equation}
and energy conservation $Q = E_e + E_\nu$.  Following the same
procedure as the electron capture case\footnote{It is perhaps less
  obvious that the assumption that the electrons are relativistic is
  justified here, given the lower integration limit.  But because
  electron phase space is only empty near or above the Fermi energy
  (and $\mu>\unit[5]{MeV}$ at the densities of interest), the
  integrand is only significant at the upper portion of the
  integration range where this approximation is justified.}
\begin{equation}
  I_{\beta} = e^{\pi\alpha Z} \int_{1}^{q}
  \frac{\epsilon^2 (\epsilon - q)^2}
  {1 + \exp[-\beta m_e c^2 (\epsilon - \mu)]} d\epsilon~.
\end{equation}
We can convert this integral into a sum of complete Fermi integrals by
making use of the mathematical identity
\begin{equation}
  \label{eq:14}
  \int_0^b \frac{x^k}{1 + \exp(x-y)} = F_k(y) - \sum_{j=0}^k \binom{k}{j} b^{k-j} F_j(y - b) ~.
\end{equation}
Defining $\vartheta \equiv \beta m_ec^2$, this yields the following expression:
\begin{equation}
\begin{split}
  \label{eq:Ibeta-fd}
  I_\beta = \frac{e^{\pi\alpha Z}}{(\beta m_e c^2)^{5}}
           \left[ F_4(\zeta-\eta) \right. & - \left. 2 \zeta F_3(\zeta-\eta) + \zeta^2 F_2(\zeta-\eta) \right] - \\
           \frac{e^{\pi\alpha Z}}{(\beta m_e c^2)^{5}}
                     \left[ F_4(\vartheta-\eta) \right. & - \\
                            F_3(\vartheta-\eta)& \times \left(4 \vartheta - 2 \zeta \right) + \\
                            F_2(\vartheta-\eta)& \times \left(6 \vartheta^2 -6 \vartheta  \zeta +\zeta ^2\right) - \\
                            F_1(\vartheta-\eta)& \times \left(4 \vartheta^3 -6 \vartheta^2 \zeta + 2\vartheta\zeta ^2\right) + \\
                            F_0(\vartheta-\eta)& \times \left(\vartheta^4 - 2\vartheta^3\zeta + \vartheta^2\zeta^2 \right) \left. \right] ~.
\end{split}
\end{equation}
Similarly, the factor $J_\beta$ necessary to calculate the neutrino
loss rate is
\begin{equation}
  \label{eq:Jbeta}
  J_{\beta} = \int_{m_ec^2}^{Q} \frac{E_e^2 E_\nu^3}{1 + \exp[-\beta (E - \mu)]} G(Z,E) dE_e~,
\end{equation}
which can be written as
\begin{equation}
\begin{split}
  \label{eq:Jbeta-fd}
  J_\beta = \frac{e^{\pi\alpha Z}}{(\beta m_e c^2)^{6}}
           \left[ F_5(\zeta-\eta) \right. & - \left. 2 \zeta F_4(\zeta-\eta) + \zeta^2 F_3(\zeta-\eta) \right] - \\
           \frac{e^{\pi\alpha Z}}{(\beta m_e c^2)^{6}}
           \left[ F_5(\vartheta-\eta) \right. & - \\
                  F_4(\vartheta-\eta)& \times \left(5 \vartheta - 3 \zeta \right) + \\
                  F_3(\vartheta-\eta)& \times \left(10 \vartheta^2 - 12 \vartheta \zeta + 3 \zeta^2 \right) - \\
                  F_2(\vartheta-\eta)& \times \left(10 \vartheta^3 - 18 \vartheta^2 \zeta + 9 \vartheta\zeta ^2 - \zeta^3 \right) + \\
                  F_1(\vartheta-\eta)& \times \left(5 \vartheta^4 - 12 \vartheta^3 \zeta + 9\vartheta^2\zeta^2 - 2
\vartheta \zeta^3 \right) - \\
                  F_0(\vartheta-\eta)& \times \left(\vartheta^5 - 3 \vartheta^4\zeta + 3 \vartheta^3\zeta^2 -\vartheta^2 \zeta^3 \right) \left. \right]~.
\end{split}
\end{equation}

Given the reaction rates and neutrino energy loss rates, we can
calculate the net heating rate of the plasma.  The energy equation for
material in the star is
\begin{equation}
  \label{eq:energy-eqn}
  T \frac{ds}{dt} = -\frac{\partial L}{\partial M} + q_* + q_\mathrm{ec} + q_\beta
\end{equation}
where $q_\mathrm{ec}$ and $q_\beta$ account for the set of weak
nuclear reactions we are considering separately and $q_*$ includes all
other heating and cooling sources such as thermal neutrino losses and
other nuclear reactions.  Under the assumption of thermal equilibrium,
the energy released by the weak reactions depends only on the total
reaction rate, the total neutrino loss rate, and the ion and electron
chemical potentials.  The energy generation rate (per capture or
decay) is
\begin{align}
  \varepsilon_\mathrm{ec} & = (-\mu_{I,Z} + \mu_{I,Z-1} + \mu_e) \lambda_\mathrm{ec} -\varepsilon_{\nu, \mathrm{ec}} \\
  \varepsilon_\beta & = (-\mu_{I,Z-1} + \mu_{I,Z} - \mu_e) \lambda_\beta - \varepsilon_{\nu, \beta}
\end{align}
where $\mu_{I,Z}$, $\mu_{I,Z-1}$ are the chemical potentials of the
ions with those charges and $\mu_e$ is the chemical potential of the
electron.  Defining $Q_g \equiv Q_{00} = (M_p - M_d)c^2$, which implicitly
making the same assumption used to derive equation \eqref{eq:q},
the specific energy generation rates are
\begin{align}
  \label{eq:qec}
  q_\mathrm{ec} & = \frac{n_\mathrm{ec}}{\rho} \varepsilon_\mathrm{ec} =  \frac{n_\mathrm{ec}}{\rho} \left[(Q_\mathrm{g} + \mu_e) \lambda_\mathrm{ec} - \varepsilon_{\nu, \mathrm{ec}}\right]~, \\
  \label{eq:qbeta}
  q_\beta & = \frac{n_\beta}{\rho} \varepsilon_\beta = \frac{n_\beta}{\rho} \left[(Q_\mathrm{g} - \mu_e) \lambda_\beta - \varepsilon_{\nu, \beta}~ \right]~,
\end{align}
where $n_\mathrm{ec}$ and $n_\beta$ are the number densities of the
species undergoing capture and decay.  Therefore, given a list of
nuclear levels and the $(ft)$-values for the transitions between them,
we can calculate the rates of and energy generation rates from
electron-capture and $\beta$-decay.

Using the above approach, it would be possible to generate tables
whose points are spaced sufficiently closely, such that interpolation
would no longer incur significant errors.  But because MESA comes with
fast quadrature routines to evaluate equation \eqref{eq:fdintegral},
it directly evaluates equations~\eqref{eq:Iec-fd}, \eqref{eq:Jec-fd},
\eqref{eq:Ibeta-fd}, and \eqref{eq:Jbeta-fd} each time one of the weak
reaction rates is needed.  While this is computationally inefficient,
the overall speed of our calculations is sufficiently unaffected that
we chose not to optimize this.

\section{Coulomb Corrections}
\label{sec:coulomb}
In a dense plasma, the electrostatic interactions of the ions and
electrons introduce corrections to the weak rates relative to rates
which assume a Fermi gas of electrons and an ideal gas of ions (as do
those presented in Appendix~\ref{sec:ecapture}).  The leading term in
the Coulomb interaction energy for ion-ion interactions
\citep[e.g.][]{Shapiro83} is
\begin{equation}
  \label{eq:Ecnaive}
  E_c = -\frac{9}{10} \frac{Z^2 e^2}{a_i}
\end{equation}
where $a_i=\left(\frac{3}{4\pi n_i}\right)^{1/3}$ is the inter-ionic
spacing.  For $Z \approx 10$, $E_c \approx -0.2 E_F$ indicating that
the interactions are energetically important.\footnote{The Coulomb
  interaction energy and the Fermi energy both scale
  $\propto \rho^{1/3}$ in the relativistic limit.}

In this section, we discuss our treatment of these corrections, and
compare our approach to previous work.  Our treatment is most similar
to that discussed in Appendix A of \citet{Juodagalvis10}.
Fig.~\ref{fig:compare_coulomb} illustrates the effects of including
these corrections on the evolution of our fiducial model.  The change
in the evolution is similar to that observed by \citet{Gutierrez96}.

\subsection{Equation of State}
\label{sec:colulomb-eos}

Most straightforwardly, a Coulomb term appears in the ion equation of
state.  This affects the weak reaction rates, because at a fixed total
pressure the electron density is lower.  The MESA equation of state
routines, which in the thermodynamic regime of interest are based on
the Helmholtz equation of state \citep{Timmes00b}, include these terms
based on the work of \citet{Yakovlev89}.

\subsection{Ion Chemical Potential}
\label{sec:colulomb-mui}

The energy required to remove an ion of one species and create an ion
of another species is given by the difference in the ion chemical
potentials.  Since electron-capture and beta-decay change the ion
charge, the presence of the Coulomb interaction energy changes the
energy difference between the parent and daughter nuclear states.  The
interaction energy is negative, and so decreasing the charge of the
nucleus (as electron-capture reactions do) requires additional energy,
which will therefore shift the onset of electron captures to higher
density.

To calculate this shift, we use the excess (that is, the part in
addition to the ideal contribution) ion chemical potential
$\mu_\mathrm{ex}$ developed in the following series of papers:
\citet{Chabrier98,Potekhin00,Potekhin09a}.  We incorporate this effect
by shifting the value of $Q$, as defined in equation~\eqref{eq:q-mu},
by an amount $\Delta E = \mu_{\mathrm{ex},p} - \mu_{\mathrm{ex},d}$.
This shift,
\begin{equation}
  \label{eq:q-shift}
  Q' = Q + \Delta E
\end{equation}
then enters the calculation of the phase space factors
and the energy generation rates.
In Fig.~\ref{fig:compare_coulomb}, the red dotted line labeled
``ion chemical potential'', shows the effect of including of these
corrections.

\subsection{Screening}
\label{sec:colulomb-vs}

\begin{figure}
  \centering
  \includegraphics[width=\columnwidth]{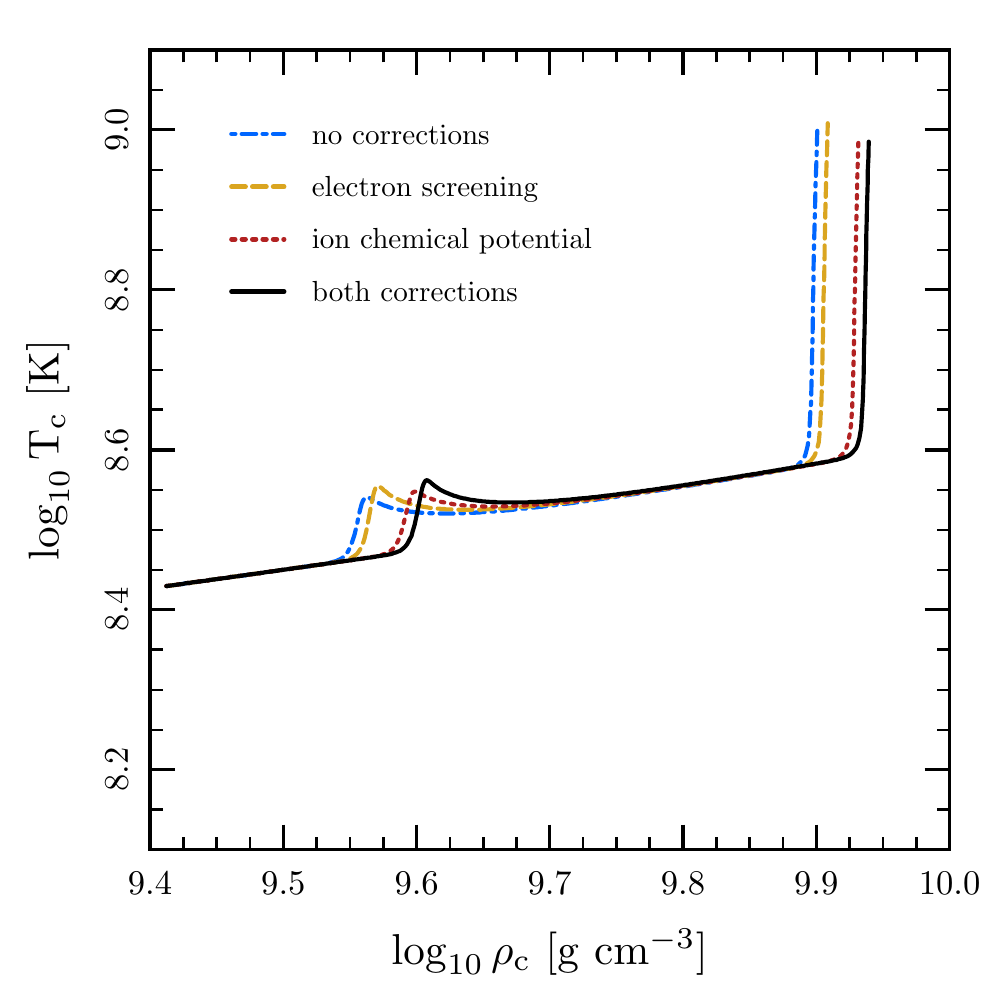}
  \caption{Illustration of the effect of Coulomb corrections on the
    evolution of the central density and temperature of the accreting
    core. (To better understand this plot the reader may first want to
    consult Fig.~\ref{fig:schematic} and the surrounding discussion.)
    All calculations include the corrections to the equation of state
    (\S~\ref{sec:colulomb-eos}).  The dashed-dotted blue line shows the result
    with no other corrections.  The dashed yellow line shows the effect
    of the inclusion of the screening corrections
    (\S~\ref{sec:colulomb-vs}).  The dotted red line shows the
    effect of the inclusion of the corrections to the ion chemical
    potential (\S~\ref{sec:colulomb-mui}).The solid black
    line shows the result with both corrections included, which is the
    default choice for our calculations.  The primary effect of
    including these corrections is a increase (of about 0.05 dex) in
    the density at which electron captures occur.}
  \label{fig:compare_coulomb}
\end{figure}

The electron density relevant to the reaction rate is not the average
electron density, but rather the electron density at the position of
the nucleus.  \citet{Itoh02} calculated the value of this screening
correction using linear response theory.  This correction can be
correctly accounted for as a shift in the value of the electron
chemical potential that enters the phase space factor.
\begin{equation}
  \label{eq:mue-shift}
  \mu'_e = \mu_e + V_s
\end{equation}
However, this correction does not enter the energy generation rates
because it has not changed the energy cost to add or remove an
electron from the bulk Fermi sea, which is the net effect of a capture
or decay.  In Fig.~\ref{fig:compare_coulomb}, the yellow dashed
line labeled ``electron screening'', shows the effect of including
these corrections.

\subsection{Comparison with Previous Work}
\label{sec:coulomb-prev}

The effect discussed in \S \ref{sec:colulomb-mui}, which is the
dominant Coulomb correction, has previously been included in studies
of ONeMg cores \citep[e.g.][]{Gutierrez96, Takahashi13}.  The approach
taken in these studies is to include this effect as a shift in the
\textit{electron} chemical potential
\begin{equation}
  \label{eq:ansatz}
  \mu'_e = \mu_e - \Delta E
\end{equation}
and to use this modified electron chemical potential in the evaluation
of the rates.  This approach is conceptually incorrect, because
$\mu_e$ and $Q$ enter the rate expression in different ways, as can be
seen in equation~\eqref{eq:Iec}.  However, given a table of
$\lambda_{ec}(\rho, T)$, one has no ability to shift $Q$, so the only
way to correct the rate is to shift the relation between $\mu_e$ and
$\rho$. In the sub-threshold case (see
equation~\ref{eq:lambda-ec-approx}), the most important term is the
exponential, which \textit{is} symmetric in $Q$ and $\mu$, and so this
approach does not lead to a substantial quantitative error in the
rate.

When making this correction both \citet{Gutierrez96} and
\citet{Takahashi13} follow \citet{Couch74} and use the form of the ion
free energy from \citet{Dewitt73}.  There has been progress in
calculating the free energy of electron-ion plasmas in the last few
decades.  As discussed in \S~\ref{sec:colulomb-mui}, we use the
fitting formula for the free energy from \citet{Potekhin09a}.  In
their work on electron capture rates in NSE material,
\citet{Juodagalvis10} use the formula quoted in \citet{Ichimaru93}.
The results of \citet{Ichimaru93} and \citet{Potekhin09a} agree, while
the shift calculated following \citet{Dewitt73} is approximately 30
per cent larger in magnitude.

The screening correction discussed in \S~\ref{sec:colulomb-vs} is not
as widely adopted.  It is included in \citet{Juodagalvis10}, but not
in \citet{Gutierrez96}.  The results of \citet{Itoh02} are within
approximately 10 per cent of the results from the Thomas-Fermi
approximation
\begin{equation}
  V_s \approx Z \sqrt{\frac{4 \alpha^3}{\pi}} E_F~.
\end{equation}
This effect has
approximately the magnitude of the difference between older ion
chemical potential and the one we adopt discussed in the preceding
paragraph, but the opposite sign.  Therefore, despite its exclusion,
the net difference between our calculations and those of
\citet{Gutierrez96} is small.

\section{Convergence}
\label{sec:convergence}

\begin{figure}
  \centering
  \includegraphics[width=\columnwidth]{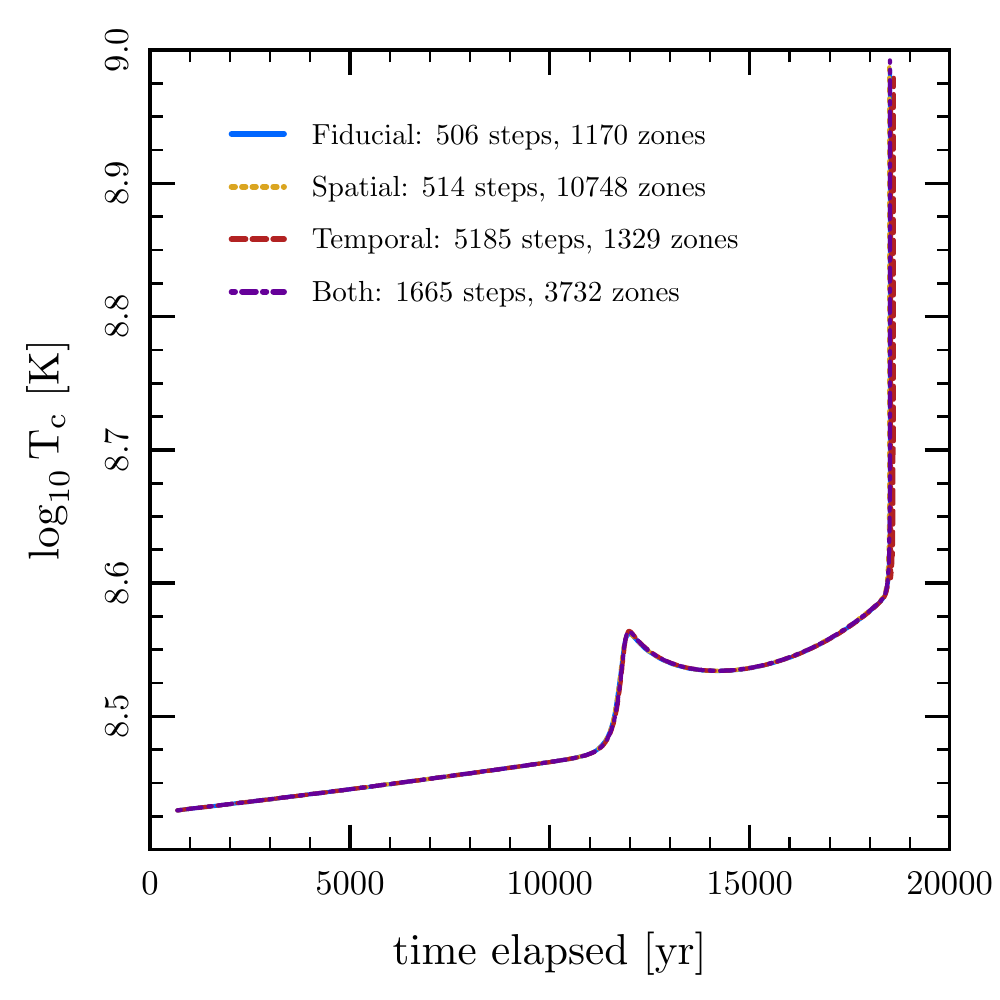}
  \caption{The legend shows the number of timesteps and the maximum
    number of zones used in the calculation of each of the runs shown
    in Table~\ref{tbl:convergence}. The negligible variation between
    models indicates our results are converged.}
  \label{fig:convergence}
\end{figure}

In order to demonstrate that our results are robust, we perform a
number of tests of the spatial and temporal convergence of our MESA
calculations.  The parameters of these runs are shown in
Table~\ref{tbl:convergence}.  We performed runs with each of the
spatial and temporal resolutions each separately ten times greater
than the fiducial case, as well as a run in which both the spatial and
temporal resolutions were three times greater than the fiducial case.
Fig.~\ref{fig:convergence} shows that the central temperature
evolution remains unchanged\footnote{The small difference at the end
  of the ``Temporal'' track is an artifact of a difference in how the
  stopping condition trigger was tripped.}and we observed that the
variation of the result in any quantity of interest was negligible.

\begin{table*}
  \begin{tabular}{lcccc}
    \hline
    Run Name & \texttt{delta\_lgRho\_cntr\_limit} & \texttt{delta\_lgRho\_cntr\_hard\_limit} & \texttt{varcontrol\_target} & \texttt{mesh\_delta\_coeff}\\
    \hline
    Fiducial & $1\times 10^{-3}$ & $3\times 10^{-3}$ & $1\times 10^{-3}$ & $1.0$ \\
    Temporal & $1\times 10^{-4}$ & $3\times 10^{-4}$ & $1\times 10^{-4}$ & -- \\
    Spatial  & -- & -- & -- & $0.1$ \\
    Both & $3\times 10^{-4}$ & $1\times 10^{-3}$ & $3\times 10^{-4}$ & $0.3$ \\
    \hline
  \end{tabular}
  \label{tbl:convergence}
  \caption{Parameters for the runs demonstrating the convergence of our results.  The column names are the specific MESA controls we used.  The controls \texttt{delta\_lgRho\_cntr\_limit}, \texttt{delta\_lgRho\_cntr\_hard\_limit}, \texttt{varcontrol\_target} control the maximal fractional change in physical variables, which in an implicit code like MESA controls the timestep.  The control \texttt{mesh\_delta\_coeff} controls the number of zones used in the calculation.  Fig.~\ref{fig:convergence} shows that central temperature evolution is essentially identical for these different runs.}
\end{table*}

In addition, we performed a run with a much larger network (203
isotopes; \texttt{mesa\_201.net} plus $\oxygen[20]$ and $\fluorine[20]$)
and confirmed that our results remained unchanged.

\section{Two-zone WD Models}
\label{sec:ztwd}

This section describes the framework we use to understand the
evolution of our MESA models after the $A=24$ captures have occurred.
The quantitative estimates shown in Figs.~\ref{fig:compression} and
\ref{fig:critical_xmg} were made using the approach described in this
Appendix.

After the $A=24$ electron captures have begun in the center of the WD,
the MESA models have two zones: an inner ``neutronized'' zone in which
the captures have occurred and $Y_e$ is lower, and an outer zone whose
composition remains unchanged.  This property of our MESA models can
be seen in Fig.~\ref{fig:final-profile}.  The neutronized zone is
growing (in a Lagrangian sense) as a function of time.

In \S~\ref{sec:two-zone} we write down an idealized model of a white
dwarf with this two zone structure.  In \S~\ref{sec:ztwd-models} we
discuss how we apply these models to understand our MESA calculations.

\subsection{Details of the Two-Zone Model}
\label{sec:two-zone}

Following \citet{Cox68}, we write down a simple model of a
zero-temperature white dwarf.  We assume spherical symmetry and
hydrostatic equilibrium and so solve the Poisson equation in spherical
coordinates:
\begin{equation}
  \label{eq:poisson}
  \frac{1}{r^2} \frac{d}{dr}\left(\frac{r^2}{\rho}\frac{dP}{dr}\right) = -4 \pi G \rho~.
\end{equation}
We also assume the equation of state of a zero temperature, ideal
Fermi gas, which is
\begin{align}
  \label{eq:zt-eos}
  P &= A f(x)\\
  x &= \left(\frac{\rho Y_e}{B}\right)^{1/3}
\end{align}
where
\begin{align}
  \label{eq:zt-eos-fab}
  f(x) &= x \sqrt{x^2+1} \left(2 x^2-3\right)+3 \sinh ^{-1}(x)\\
\nonumber
  A &= \frac{\pi m_e^4 c^5}{3 h^3} \approx \unit[6.0 \times 10^{22}]{dynes\,cm^{-2}}\\ \nonumber
  B &= \frac{8 \pi m_e^3 c^3}{3 h^3 N_A} \approx \unit[9.7 \times 10^5]{g\,cm^{-3}}~.
\end{align}
Combining this equation of state with equation~\eqref{eq:poisson}
gives
\begin{equation}
  \label{eq:5}
  \frac{1}{r^2} \frac{d}{dr}\left[r^2 \frac{d}{dr}\left(x^2 + 1\right)^{1/2}\right] = -\frac{\pi G B^2}{2 A Y_e^2} x^3~\,
\end{equation}
where we have made the assumption that $dY_e/dr = 0$.

In order to non-dimensionalize these equations, define $z^2 = x^2 + 1$
and let $z_c$ be the value of $z$ at center of the model.  We also define
\begin{equation}
  \alpha \equiv \left(\frac{2 A}{\pi G}\right)^{1/2} \frac{1}{B z_c} ~.
\end{equation}
and transform to the variables
\begin{align}
  r &\equiv \alpha \zeta \\
  z &\equiv z_c \Phi~.
\end{align}
This yields the differential equation
\begin{equation}
  \label{eq:ztwd-final}
  \frac{d}{d\zeta}\left(\zeta^2 \frac{d\Phi}{d\zeta}\right) = \frac{1}{Y_e^2} \left(\Phi^2-\frac{1}{z_c^2}\right)^{3/2}~.
\end{equation}
At the center ($\zeta=0$), the boundary conditions are
\begin{align}
  \label{eq:bc-center}
  \Phi(\zeta = 0) &= 1 \\
  \left.\frac{d\Phi}{d \zeta}\right|_{\zeta=0} &= 0
\end{align}
At the surface ($\zeta=\zeta_s$), $\rho \to 0$ and so $z \to 1$,
meaning
\begin{equation}
  \label{eq:bc-surface}
  \Phi(\zeta = \zeta_s) = \frac{1}{z_c}~.
\end{equation}

Now, we divide the white dwarf into two zones which have different
values of $Y_e$.  By assuming a piecewise constant form for $Y_e$, we
can continue to solve equation~\eqref{eq:ztwd-final} throughout the
whole white dwarf. Specifically, we use
\begin{equation}
  Y_e =
  \begin{cases}
      Y_{e,0} & \text{if }z < z_n \\
      Y_{e,n} & \text{if }z > z_n~.
  \end{cases}
  \label{eq:eos-neut}
\end{equation}
The transition between the two zones occurs at $\zeta_n$ such that
$\Phi(\zeta_n) = z_n/z_c$. The following physical conditions must be
satisfied at this interface
\begin{align}
  \label{eq:bc-physical}
  P_+ &= P_- \\
  \left(\frac{1}{\rho}\frac{dP}{dr}\right)_+&= \left(\frac{G M_r}{r^2}\right)_{-}~.
\end{align}
Note that the continuity of $P$ implies the continuity of $x$, and
hence $z$, even though $Y_e$ is discontinuous.  The dimensionless
equivalents of these conditions are
\begin{align}
  \label{eq:bc-dimensionless}
  \Phi(\zeta = \zeta_n^-) &= \Phi(\zeta = \zeta_n^+) \\
  \left(Y_{e}\frac{d\Phi}{d \zeta}\right)_{\zeta=\zeta_n^-}&= \left( Y_{e} \frac{d\Phi}{d \zeta}\right)_{\zeta=\zeta_n^+}~.
\end{align}

Constructing a two-zone model is now simple.  Specify the three
parameters for the equation of state: $Y_{e,0}, Y_{e,n}, z_n$.  Select
a central density (which sets the value of $z_c$) and then integrate
the ODE observing the boundary and jump conditions. The solution gives
the structure of a single two-zone model.  A one-parameter family of
models can be constructed by varying $z_c$, which in turn varies the
properties (e.g., mass, radius) of the model.

\subsection{Applications of the Two-Zone Model}
\label{sec:ztwd-models}

Our MESA models are in hydrostatic equilibrium: their evolution is
occurring on timescales much longer than the dynamical time.  We use
the two-zone model to find approximate sequences of hydrostatic models
along which the MESA models evolve. This gives us insight into the
processes that control the timescale of the evolution.

The piecewise equation of state given in equation~\eqref{eq:eos-neut}
can be used to represent the $A=24$ captures: setting
$\log_{10} \rho_n = 9.6$, with $Y_{e,0} = 0.5$ and
$Y_{e,n} = Y_{e,0} - X_\mathrm{Mg} / 12$ corresponds to instantaneous
neutronization of all available $\magnesium$ at densities above the
threshold density.

\begin{figure}
  \centering
  \includegraphics[width=\columnwidth]{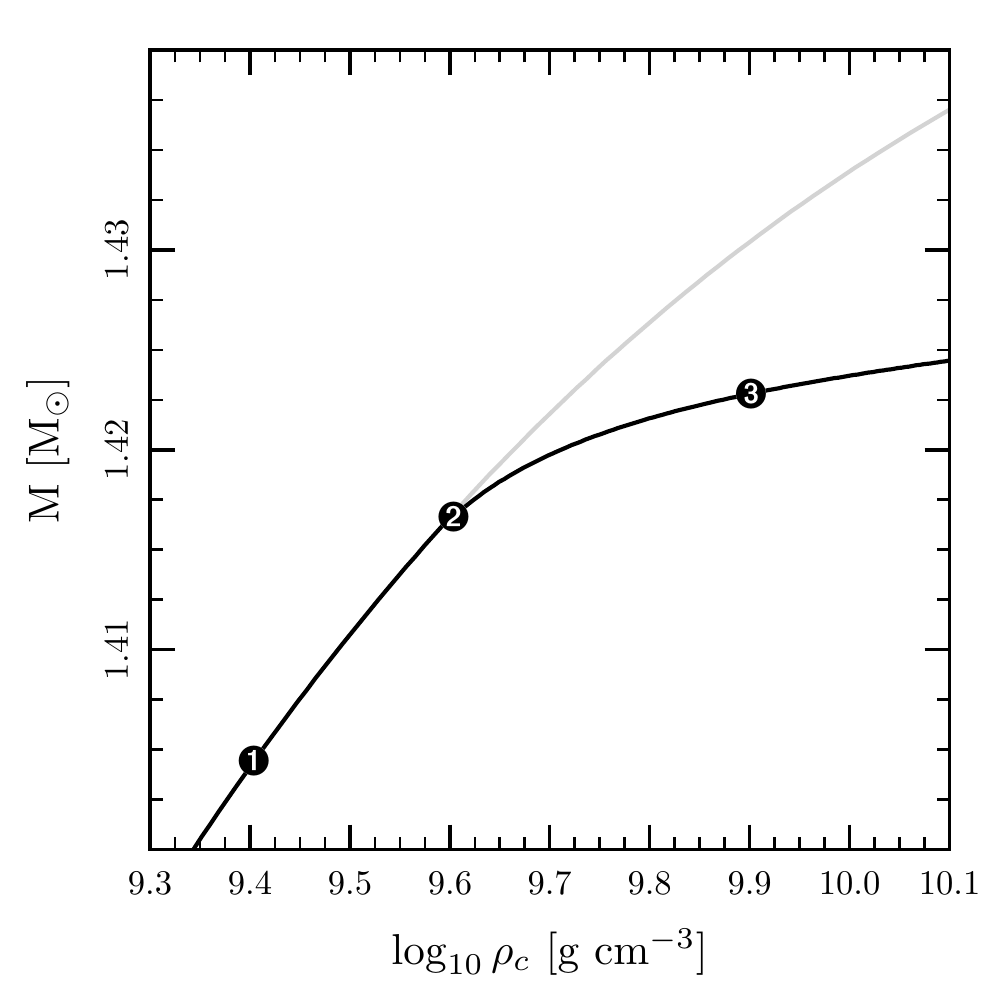}
  \caption{The black line shows the sequence of two-zone hydrostatic
    models with electron captures above $\log_{10} \rho_n = 9.6$ with
    the change in $Y_e$ corresponding to $X_\mathrm{Mg} = 0.05$.  The
    grey line shows a zero-temperature model without neutronization.
    With neutronization taken into account, the central density
    increases more rapidly with increasing mass.  The numbered points
    indicate a temporal series of models; specific points are
    discussed in more detail in the text.}
  \label{fig:ztwd2-xmg-0p05}
\end{figure}

\begin{figure}
  \centering
  \includegraphics[width=\columnwidth]{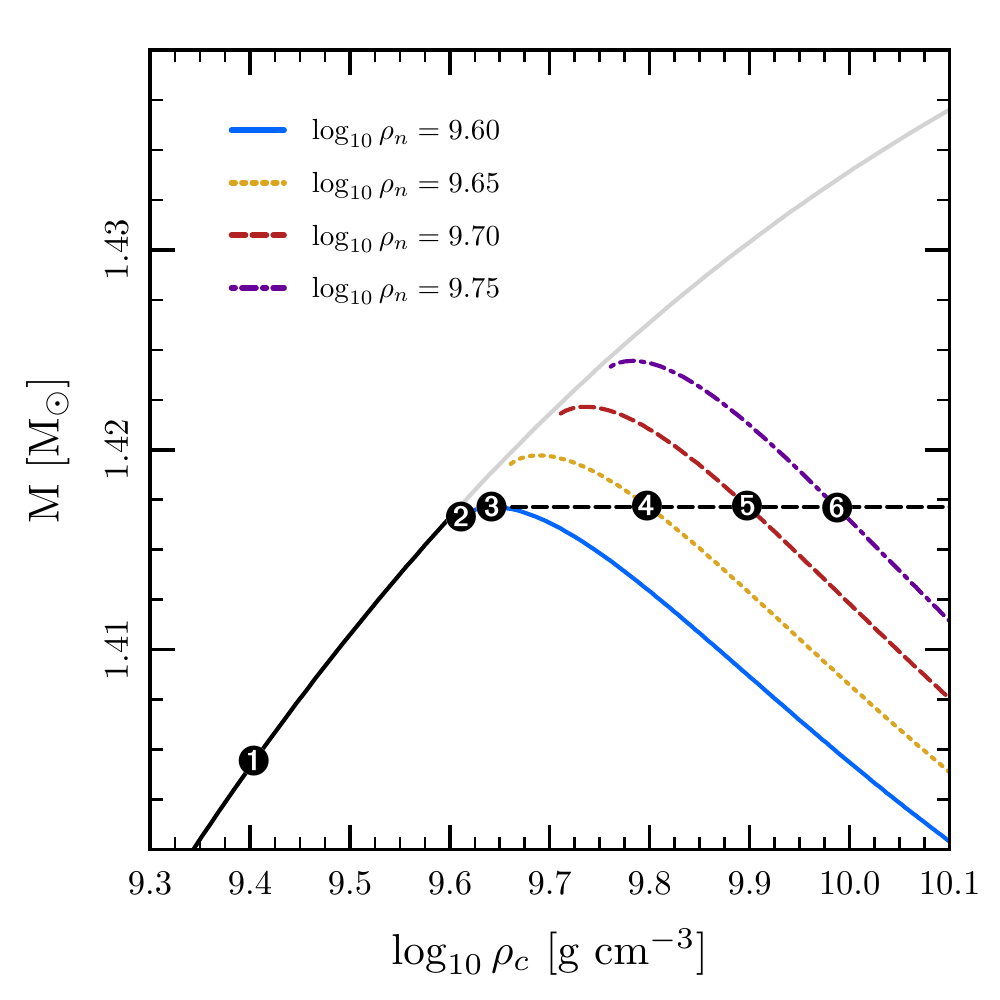}
  \caption{The black line shows the sequence of two-zone hydrostatic
    models corresponding to $X_\mathrm{Mg} = 0.15$.  The grey line
    shows a zero-temperature model without neutronization.  The
    colored lines show models with neutronization at different
    densities.  The gaps between the colored lines and the grey line
    are the discontinuities in $\rho_c$ caused by the discontinuity in
    $Y_e$; the sequence in continuous in $P_c$.  The numbered points
    indicate a temporal series of models; specific points are
    discussed in more detail in the text.  The black line changes from
    solid to dashed when the sequence of models changes from being
    defined by constant $\rho_n$ to constant $M$.}
  \label{fig:ztwd2-xmg-0p15}
\end{figure}

Fig.~\ref{fig:ztwd2-xmg-0p05} shows the schematic evolution of models
with $X_\mathrm{Mg} = 0.05$.  The black line is the family of two-zone
hydrostatic models.  This family of models is generated by varying
$P_c$ (the central pressure). Because of the discontinuity in $Y_e$,
the continuous variation in $P_c$ gives a discontinuous variation in
$\rho_c$.  (In Fig.~\ref{fig:ztwd2-xmg-0p05}, the density jump is
hidden by point 2, but the jumps are apparent in
Fig.~\ref{fig:ztwd2-xmg-0p15}.) The grey line shows the family of
models without neutronization.  For central densities less than
$\rho_n$ (e.g., point 1) the two families are equivalent.  At point 2,
the central density crosses the threshold density and the families
diverge.  At point 3, the model has a substantial low-$Y_e$ core and
has a much higher central density at fixed mass relative to the models
without neutronization.

The numbered points in Fig.~\ref{fig:ztwd2-xmg-0p05} represent a
temporal sequence of models.  The time evolution is driven by the
increase in $M$ set by accretion.  For a given value of $\dot{M}$, the
timescale for changes in $\rho_c$ is given by
equation~\ref{eq:tcompress}, where the value of $d\ln \rho_c/d \ln M$
comes from the sequence of hydrostatic models.  In
Fig.~\ref{fig:compression}, the black dashed line is calculated using
the model without neutronization, while the black dotted line is
calculated using the two zone model.  This latter line does an
excellent job of quantitatively describing the more rapid contraction
of the MESA model following the $A=24$ captures.

Fig.~\ref{fig:ztwd2-xmg-0p15} shows the schematic evolution of models
with $X_\mathrm{Mg} = 0.15$. The black line is the family of two-zone
hydrostatic models.  The grey line shows the family of models without
neutronization.  For central densities less than $\rho_n$ (e.g., point
1) the two families are equivalent.  At point 2, neutronization begins
to have an effect.  At point 3, the model passes the maximum mass
possible for the family of models with $\log_{10} \rho_n = 9.60$.  Up
to this point, as in the case with $X_\mathrm{Mg} = 0.05$, the points
1-3 represent a temporal sequence of models whose time evolution is
driven by increasing $M$.

With the equation of state held fixed, point 3 would mark the onset of
dynamical instability.  However, the characteristic timescale of the
electron captures is longer than the dynamical time.  Therefore it is
not physically possible for $\rho_n$ to remain fixed.  Only material
at densities where the electron capture timescale is shorter than the
evolutionary timescale is able to completely neutronize.

For the large $X_\mathrm{Mg}$ models like that in
Fig.~\ref{fig:ztwd2-xmg-0p15}, the evolutionary timescale becomes
sufficiently short that there is no longer time for a significant
amount of mass to accrete.  Therefore, the evolution switches to a
sequence of models with constant $M$, as indicated by the black dashed
line and points 4-6 in Fig.~\ref{fig:ztwd2-xmg-0p15}. The fixed value
of $M$ is the maximum mass for a model with the initial value of
$\log_{10} \rho_n = 9.60$ (i.e., point 3).

The evolution along the temporal sequence of points 3-6 is limited by
the electron capture rates.  To describe this quantitatively, for a
given $\rho_c$, we find the value of $\rho_n$ that corresponds to the
hydrostatic model with a given $M$.  Then we calculate the
neutronization timescale (equation~\ref{eq:tneut-mg}) corresponding to
this value of $\rho_n$.  This gives the black dash dotted line shown
in Fig.~\ref{fig:compression}, which does a good job of reproducing
the evolution observed in the MESA model.  For simplicity, we used the
electron capture rates at fixed temperature ($\log_{10} T = 8.6$) to
calculate $\lambda_\mathrm{ec}$.

The qualitatively different evolution experienced by the
$X_\mathrm{Mg} = 0.05$ and $X_\mathrm{Mg} = 0.15$ models is due to the
presence of a maximum mass in the families of two-zone models at a
central density less than the critical density for $\neon$ captures.
Calculating the central density at which this maximum occurs for each
value of $X_\mathrm{Mg}$ gives the critical curve shown as a black
dashed line in Fig.~\ref{fig:critical_xmg}.

\end{document}